\documentclass[11pt]{article}
\usepackage{amsfonts}
\usepackage{amsmath,amssymb,amscd}
\usepackage{stmaryrd}
\usepackage[dvips]{color}
\usepackage[dvipdfmx]{hyperref}
\usepackage{fancybox}
\usepackage{cite}
\usepackage{bm}
\usepackage[all]{xy}
\usepackage{wrapfig}
\usepackage{ulem}   
\usepackage{braket}

\baselineskip=\normalbaselineskip

\begin{document} 

\numberwithin{equation}{section}
\renewcommand{\theequation}{\thesection.\arabic{equation}}
%%%%% Begin the Note, use \section{} %%%%%%%%%
\def\natural{\mathbb{N}}
\def\mat#1{\matt[#1]}
\def\matt[#1,#2,#3,#4]{\left(%
\begin{array}{cc} #1 & #2 \\ #3 & #4 \end{array} \right)}

%%%%%%special for this file only %%%%%%%%%%%%
\def\bea#1\ena{\begin{align}#1\end{align}}
\def\bean#1\enan{\begin{align*}#1\end{align*}}
\def\p{\partial}
\def\nn{\nonumber\\}
\def\cL{{\cal L}}

\null \hfill Preprint: TU-1050  \\[3em]
\begin{center}
{\LARGE \bf{

Contravariant geometry and emergent gravity from noncommutative gauge theories
}}\\[2em] 
\end{center}

\begin{center}
{Yukio Kaneko${}^{\sharp}$\footnote{
e-mail: y\_kaneko@tuhep.phys.tohoku.ac.jp}},
{Hisayoshi Muraki${}^{\flat}$\footnote{
e-mail: hmuraki@sogang.ac.kr
}}
and {Satoshi Watamura${}^{\sharp}$\footnote{
e-mail: watamura@tuhep.phys.tohoku.ac.jp}}\\[2em] 

${}^{\sharp}$ Graduate School of Science,\\
Tohoku University,\\
Aoba-ku, Sendai 980-8578, Japan\\[1em] 

${}^{\flat}$ Department of Physics,\\
Sogang University,\\
Seoul 121-742, Korea\\[5em]

\thispagestyle{empty}

\abstract{
\noindent
We investigate a relation of the contravariant geometry to the emergent gravity from noncommutative gauge theories.
We give a refined formulation of the contravariant gravity and provide solutions to the contravariant Einstein equation.
We linearize the equation around background solutions, including curved ones.
A noncommutative gauge theory on the Moyal plane can be rewritten as 
an ordinary gauge theory on a curved background via the Seiberg-Witten map, which is known as the emergent gravity.
We show that this phenomenon also occurs 
for a gauge theory on a noncommutative homogeneous K\"ahler background.
We argue that the resulting geometry can be naturally described by the contravariant geometry 
under an identification of the fluctuation of the Poisson tensor with the field strength obtained by the Seiberg-Witten map. 
These results indicate that the contravariant gravity is a suitable framework for noncommutative spacetime physics.
} 

\end{center}

\vskip 2cm

\eject
%%%%%%%%%%%%%%%%%%%%%%%%%%%%%%%%%%%%%%%%%%%%%%%

\tableofcontents

%%%%%%%%%%%%%%%%%%%%%%%%%%%%%%%%%%%%%%%%%%%%%%%
\section{Introduction}

The theory of quantum gravity is a long standing problem.
At present, string theory is one of the promising approaches to this problem.
In string theory, the string length $l_{s}$ can be considered as a natural cut-off scale,
which gives a minimal length in the low energy physics.
This idea is implemented by introducing an uncertainty in spacetime \cite{Yoneya}.
In analogy to quantum mechanics, where the uncertainty on phase space implies noncommutativity of 
phase space coordinates, the uncertainty of spacetime implies noncommutativity of spacetime coordinates.

In quantum mechanics, the noncommutativity of observables can be obtained by 
the quantization of the corresponding Poisson structure.
Conversely, the Poisson structure can be derived from the quantum system by a semi-classical approximation.
Analogously, the noncommutativity of spacetime would be obtained by the quantization of a certain Poisson structure.
The problem is that gravity on such a noncommutative spacetime is not well understood.

The contravariant gravity provides one approach to this problem.
Since the contravariant gravity contains both a metric and a Poisson tensor as dynamical variables,
it can be considered as a semiclassical approximation of gravitational physics on a noncommutative spacetime \cite{KMW}.
This theory is based on the contravariant connection \cite{Vaisman1,Vaisman2,fernandes2000,Boucetta:2011ofa2,Boucetta:2011ofa3,KSmodular,Hawkins:2002rf} that is determined by compatibility of the Riemann structure 
and the Poisson structure \cite{AMW}.
This kind of gravity theories has been also discussed to understand the nongeometric background 
and the nongeometric fluxes which appear after performing successive T-duality transformations 
in string theory and supergravity \cite{Andriot:2011uh, Andriot:20142, AMW, AMSW, Blumenhagen:2012nt,Andriot:2012wx}.
In the framework of double field theory, such gravity theories
appear as well \cite{Andriot:2012an,Andriot:2013xca,Andriot:20141}.

Gauge theories on a noncommutative spacetime, 
noncommutative gauge theories for short, are better understood than gravity on a noncommutative spacetime.
In string theory, the noncommutative gauge theory with the Moyal star product appears as an effective theory on 
a stack of D-branes on a constant NSNS B-field background \cite{SeibergWitten}.
In the paper \cite{SeibergWitten}, a notion of the Seiberg-Witten map is formulated, 
which gives a relation between noncommutative gauge theories and usual gauge theories.

Some noncommutative field theories appear as a certain continuum limit 
of the matrix theories which are expected as constructive definitions of string theory and quantum gravity \cite{BFSS,IKKT}.
Once the classical solution of the matrix theory is given, 
one can find the corresponding spacetime with a metric and some other geometric structures \cite{Stein,I,IMM}.
In terms of the matrix theory, the star product is derived by the continuum limit of the matrix product, 
and the gauge field can be identified with the fluctuation around the solution of the matrix theory.
For instance, the star product on the complex projective planes can be realized in the large matrix size limit 
of the matrix product  \cite{BDLMO}.
From the BFSS matrix model, the IKKT matrix model and their variants,
one can obtain the noncommutative gauge theory on the Moyal plane \cite{AIIKKT} and 
that on the fuzzy tori \cite{CDS} and the fuzzy sphere \cite{IKTW}. See also \cite{watamurawatamura}.

An interesting relation between gravity and noncommutative gauge theory on the Moyal plane is pointed out in \cite{Rivelles,Yang}.
A gauge theory on a noncommutative flat space can be re written by a nonlinearly interacting gauge theory
on a curved background via the Seiberg-Witten map.
The curved background obtained in this procedure is induced by the gauge field. 
The background curvature becomes nonzero if the gauge field strength is nontrivial,
even though the original noncommutative space is flat.
Since this can be interpreted that a geometric structure emerges from gauge theories on noncommutative space, 
this phenomenon is called emergent gravity or emergent geometry \cite{Rivelles,Yang,Steinacker,Steinacker3,YS,Y}.

The main focus of this paper is to provide a geometric picture of the metric arising in the emergent gravity.
We argue that the contravariant geometry gives an appropriate geometric framework to describe the emergent gravity. 
We find that the corresponding curvature has a simple interpretation as the curvature in the contravariant gravity.
This is also the case when we start with a noncommutative gauge theory on a curved background.
To be more specific, we consider a noncommutative gauge theory 
on a homogeneous K\"ahler manifold by using an action which is constructed in \cite{MSSU,Sako}.

\vspace{5mm}

Paper is organized as follows. In section 2, we give a refined formulation of the contravariant gravity 
which is based on the Poisson-Lie algebroid.
We provide the equation of motion, a corresponding action and gauge symmetry structure of the contravariant gravity.
Examples of solutions of the equation of motion are presented.
We also consider the linearization of the equation of motion.
In this paper, the fluctuation of the Poisson tensor is introduced in addition to the fluctuation of the metric.
We give a general formula of the linear approximation of the contravariant Riemann tensor around a general background.

In section 3, we discuss the relation between the emergent gravity and the contravariant geometry.
Following the argument by \cite{Rivelles}, we first review how to relate gravity with noncommutative gauge theories.
For the Moyal plane case, we find that the Ricci scalar obtained in emergent gravity 
coincides with the contravariant Ricci scalar constructed by the flat metric and a certain configuration of the Poisson tensor.
The fluctuation of the Poisson tensor turns out to be identified with that of the gauge field.

In section 4, we first apply the emergent gravity argument to a gauge theory on a noncommutative
homogeneous K\"ahler manifold \cite{MSSU,Sako}. 
We find that the resulting geometry does not always allow a K\"ahler structure.
We present a condition on the gauge field for the emergent metric to satisfy the K\"ahler condition.
In this case, we find that the contravariant geometry works also  as an appropriate framework,
where we simply assume that the background Poisson tensor is given by the symplectic structure 
of the background K\"ahler manifold.
The fluctuation of the Poisson tensor of the contravariant gravity can be identified with
the fluctuation of the gauge field strength.
Under this identification we find that the contravariant Ricci scalar coincides with 
the ordinary Ricci scalar of the geometry obtained in the emergent gravity.

%%%%%%%%%%%%%%%%%%%%%%%%%%%%%%%%%%%%%%%%%%%%
%%%%%%%%%%%%%%%%%%%%%%%%%%%%%%%%%%%%%%%%%%%%
\section{Contravariant gravity}

In this section, we revisit the formulation of the contravariant gravity.
Fundamental materials are already given in \cite{AMW} and also \cite{KMW}.
Here, we explain briefly the construction of the contravariant gravity and
its gauge symmetry. 
Let  $(M,G,\pi)$ be a Riemann-Poisson manifold, where $M$ is a smooth manifold, $G^{ij}$ is a metric and 
$\pi=(1/2)\pi^{ij}\partial_{i}\wedge\partial_{j}$ is a Poisson bi-vector.
The equation of motion of the contravariant gravity without matter fields is given by
\begin{align}
\bar{R}^{ij}-\frac{1}{2}G^{ij}\bar{R}=0,		\label{eom1}
\end{align}
where $\bar{R}$ is an analog of the Ricci scaler which will be given in (\ref{eq:Riccis}).
We call this equation the contravariant Einstein equation.
The corresponding action will turn out to be invariant under the diffeomorphism
with a gauge parameter $X^{i}$ which is given by
\begin{align}
\delta_{X} G^{ij}
	=&X^{k}\partial_{k}G^{ij}-G^{il}\partial_{l}X^{j}-G^{jl}\partial_{l}X^{i},\\
\delta_{X} \pi^{ij}
	=&X^{k}\partial_{k}\pi^{ij}-\pi^{il}\partial_{l}X^{j}+\pi^{jl}\partial_{l}X^{i},
\end{align}
and also the  
$\beta$-diffeomorphism \cite{Blumenhagen:2012nt} with a gauge parameter $\zeta_{i}$ which is given by
\begin{align}
\bar{\delta}_{\zeta} G^{ij}
	=&\zeta_{k}\pi^{kl}\partial_{l}G^{ij}
	+(\pi^{ik}\partial_{k}\zeta_{l}+\zeta_{k}\partial_{l}\pi^{ik})G^{lj}
	+(\pi^{jk}\partial_{k}\zeta_{l}+\zeta_{k}\partial_{l}\pi^{jk})G^{il},\\
\bar{\delta}_{\zeta} \pi^{ij}
=&0,
\end{align}
where we require that the gauge transformation does not break the Poisson condition.
In the following subsections, after introducing the Poisson-Lie algebroid and the Poisson-Riemann geometry,
we formulate an action of the contravariant gravity theory and see its gauge invariance.
Finally, we provide solutions of the contravariant Einstein equation and discuss the linearized contravariant gravity.

%%%%%%%%%%%%%%%%%%%%%%%%%%%%%%%%%%%%%%%%%%%%
\subsection{Poisson-Lie algebroid}

We give a definition of the Poisson-Lie algebroid $(T^{*}M,[\cdot,\cdot]_{\pi},\rho)$.
Here the bracket $[\cdot,\cdot]_{\pi}:\Gamma(T^{*}M)\times \Gamma(T^{*}M)\rightarrow\Gamma(T^{*}M)$ 
is the Koszul bracket given by
\begin{align}
	[\xi,\eta]_{\pi}
	=\left(\xi_{k}\pi^{kl}\partial_{l}\eta_{i}
	-(\pi^{lk}\partial_{k}\xi_{i}
	+\partial_{i}\pi^{lk}\xi_{k})\eta_{l} \right)dx^{i},
\end{align}
for 1-forms $\xi$ and $\eta$.
A map $\rho:\Gamma(T^{*}M)\rightarrow \Gamma(TM)$ is the anchor map defined by
\begin{align}
\rho(\xi)=\xi_{i}\pi^{ij}\partial_{j},
\end{align}
where $\xi$ is a 1-form\footnote{Hereafter 
the set of sections of the cotangent bundle $T^*M$ shall be denoted by $T^*M$ for notational simplicity,
as far as there is no potential for confusing.}.
The triple $(T^{*}M,[\cdot,\cdot]_{\pi},\rho)$ is called the Poisson-Lie algebroid which satisfies axioms of the Lie algebroid:
\begin{align}
&\rho([\xi,\eta]_{\pi})
=
[\rho(\xi),\rho(\eta)],\\
&[\xi,f\eta]_{\pi}
=
[\xi,\eta]_{\pi}+(\rho(\xi)f)\eta,
\end{align}
where $f$ is a function on $M$.

We give a geometrical understanding of the Poisson-Lie algebroid which will play an important role 
in the discussion on the invariance of the contravariant gravity.
In usual differential geometry, a Lie bracket between vector fields $X,Y$ can be written by a Lie derivative $[X,Y]={\cal L}_{X}Y$.
We can also interpret the Koszul bracket as a Poisson-Lie derivative of 1-forms $[\xi,\eta]_{\pi}=:\bar{\cal L}_{\xi}\eta$ which is extended for general tensors in the following way.
First, the Schouten-Nijenhuis bracket for an $n$-vector ${\cal X}=X_{1}\wedge\cdots \wedge X_{n}$ and an $m$-vector ${\cal Y}=Y_{1}\wedge\cdots \wedge Y_{m}$ is 
\begin{align}
[{\cal X},{\cal Y}]_{\text{SN}}=\sum_{i,j}(-1)^{i+j}[X_{i},Y_{j}]\wedge X_{1}\wedge\cdots\wedge X_{i-1}\wedge X_{i+1}\wedge\cdots\wedge X_{n}\nonumber\\
\wedge Y_{1}\wedge\cdots\wedge Y_{j-1}\wedge Y_{j+1}\wedge\cdots\wedge Y_{m},
\end{align}
where the bracket in the right hand side is the ordinary Lie bracket between the vector fields.
The Lichnerowicz differential $d_{\pi}:\wedge^{n}TM\rightarrow\wedge^{n+1}TM$ is defined by
\begin{align}
d_{\pi}V=[\pi,V]_{\text{SN}},
\end{align}
where $V=(1/n!)V^{i_{1}\cdots i_{n}}\partial_{i_{1}}\wedge\cdots \wedge\partial_{i_{n}}$ is an element of $\wedge^{n}TM$.
This  derivative is nilpotent $d_{\pi}^2=0$ if and only if the bi-vector $\pi$ satisfies the Poisson condition, $[\pi,\pi]_{\text{SN}}=0$.
An interior product $\bar{\iota}_{\xi}:\wedge^{n}TM\rightarrow\wedge^{n-1}TM$ is given by
\begin{align}
\bar{\iota}_{\xi}V=\frac{1}{(n-1)}\xi_{i}V^{ij_{1}\cdots j_{n-1}}\partial_{j_{1}}\wedge\cdots \wedge\partial_{j_{n-1}},
\end{align}
where $\xi=\xi_{i}dx^{i}$ is a 1-form.
The Poisson-Lie derivative is defined for poly-vectors
\begin{align}
\bar{\cal L}_{\xi}=d_{\pi}\bar{\iota}_{\xi}+\bar{\iota}_{\xi}d_{\pi},\label{ualgfnxhlfbjhxdfngfbd1}
\end{align}
for a function $f$
\begin{align}
\bar{\cal L}_{\xi}f=\pi^{ij}\partial_{j}f\partial_{i},\label{ualgfnxhlfbjhxdfngfbd2}
\end{align}
which satisfies
\begin{align}
[\bar{\cal L}_{\xi},\bar{\cal L}_{\eta}]=&\bar{\cal L}_{[\xi,\eta]_{\pi}},\\
\bar{\cal L}_{\xi}(fV)=&\bar{\cal L}_{\xi}f\wedge V+f\bar{\cal L}_{\xi}V,
\end{align}
for any vector field $V$.
For an 1-form the Poisson-Lie derivative is $\bar{\cal L}_{\xi}\eta=[\xi,\eta]_{\pi}$, which is consistent with (\ref{ualgfnxhlfbjhxdfngfbd1}) and (\ref{ualgfnxhlfbjhxdfngfbd2}).
For the Poisson-Lie derivative for a poly-form $\alpha=\alpha_{1}\wedge\cdots \wedge \alpha_{n}$ is defined by
\begin{align}
\bar{\cal L}_{\xi}\alpha=\sum_{i=1}^{n}(-1)^{i-1}\bar{\cal L}_{\xi}\alpha_{i}\wedge\alpha_{1}\wedge\cdots \wedge\alpha_{i-1}\wedge\alpha_{i+1}\wedge\cdots\wedge \alpha_{n}.
\end{align}
A set of operations $(d_{\pi},\bar{\cal L},\bar{\iota})$ satisfies the Cartan algebra.

Finally, we comment on a ``$\beta$-isometry" of a Poisson manifold $(M,\pi)$ based on the Poisson-Lie algebroid.
The isometry is generated by the Poisson-Lie derivative for a 1-form $\zeta$ such that
\begin{align}
&\bar{\cal L}_{\zeta}[\xi,\eta]_{\pi}=[\bar{\cal L}_{\zeta}\xi,\eta]_{\pi}+[\xi,\bar{\cal L}_{\xi}\eta]_{\pi},\\
&\rho(\bar{\cal L}_{\zeta}\xi)=\bar{\cal L}_{\zeta}\rho(\xi),
\end{align}
for any $\eta, \zeta \in T^{*}M$.
This conditions are equivalent with
\begin{align}
\bar{\cal L}_{\zeta}\pi^{ij}=0.\label{eq:symcon}
\end{align}
This isometry group is an subset of the $\beta$-diffeomorphism in the contravariant gravity, which we will explain later.

%%%%%%%%%%%%%%%%%%%%%%%%%%%%%%%%%%%%%%%%%%%%%%%
\subsection{Poisson-Riemann geometry}
In this subsection, we introduce a notion of the connection for the Poisson-Lie algebroid.
We can define geometric quantities for the connection: a Riemann curvature and a torsion tensor, as usual geometry for a connection.
This geometry is also studied in mathematical contexts \cite{fernandes2000,Boucetta:2011ofa2,Boucetta:2011ofa3,KSmodular}.

An affine connection of the Poisson-Riemann geometry is a map $\bar{\nabla}:T^{*}M\times T^{*}M\rightarrow T^{*}M$ which satisfies the following axioms
\bea
	&\bar\nabla_{\xi}(\eta+\zeta)=\bar\nabla_{\xi}\eta+\bar\nabla_{\xi}\zeta,\\
	&\bar\nabla_{\xi+\eta}\zeta=\bar\nabla_{\xi}\zeta+\bar\nabla_{\eta}\zeta,\\
	&\bar\nabla_{f\xi} \eta=f \bar\nabla_\xi \eta, \\
	&\bar\nabla_\xi (f\eta)=(\bar\cL_{\xi} f)\eta+f \bar\nabla_{\xi} \eta , \label{affine}
\ena
where $\xi, \eta$ and $\zeta$ are 1-forms.
The curvature and the torsion of the affine connection are defined by
\begin{align}
&\bar{R}(\xi,\eta)\zeta=\bar{\nabla}_{\xi}\bar{\nabla}_{\eta}\zeta-\bar{\nabla}_{\eta}\bar{\nabla}_{\xi}\zeta-\bar{\nabla}_{[\xi,\eta]_{\pi}}\zeta,\\
&\bar{T}(\xi,\eta)=\bar{\nabla}_{\xi}\eta-\bar{\nabla}_{\eta}\xi-[\xi,\eta]_{\pi}.
\end{align}
These quantities are covariant, since they satisfy
\begin{align}
&\bar{R}(f\xi,g\eta)h\zeta =fgh\bar{R}(f,g)\zeta,\\
&\bar{T}(f\xi,g\eta)=fg\bar{T}(\xi,\eta),
\end{align}
for functions $f, g$ and $h$ on $M$.

In terms of a local coordinate $\{x^i\}$, the connection is specified by coefficients
\bea
	\bar\nabla_{dx^i}dx^j = \bar{\Gamma}^{ij}_k dx^k, \label{25}
\ena
for 1-form basis $\{dx^i\}$.
Then, the curvature tensor and the torsion tensor can be expressed as
\begin{align}
&\bar{T}(dx^{i},dx^{j})=\bar{T}^{ij}_{m}dx^{m}=(\bar{\Gamma}^{ij}_{m}-\bar{\Gamma}^{ji}_{m}-\partial_{m}\pi^{ij})dx^{m},\label{eq:T}\\
&\bar{R}(dx^{i},dx^{j})dx^{k}=\bar{R}^{kij}_{l}dx^{l}= (\pi^{im} \p_m \bar{\Gamma}^{jk}_l 
		- \pi^{jm} \p_m \bar{\Gamma}^{ik}_l -\p_n \pi^{ij} \bar{\Gamma}^{nk}_l 
		+ \bar{\Gamma}^{jk}_m \bar{\Gamma}^{im}_l 
		 - \bar{\Gamma}^{ik}_m \bar{\Gamma}^{jm}_l )dx^{l}.	\label{eq:R}
\end{align}

%%%%%%%%%%%%%%%%%%%%%%%%%%%%%%%%%%%%%%%%%%%%
\subsection{Contravariant gravity}

In this subsection, we review formulation of the contravariant gravity.
So far we have considered the fixed Poisson tensor in the Poisson-Lie algebroid.
From now on, we consider both the metric and the Poisson tensor as variables of the physical theory.

We introduce a derivative by using the Poisson tensor:
\begin{align}
\{x^{i},\cdot\}_{\text{PB}}=\pi^{ij}(x)\partial_{j},\label{eringi}
\end{align}
and its covariantization of this derivative under diffeomorphism by using a connection $\bar{\Gamma}$ in the following way,
\begin{align}
\bar{\nabla}^{i}V^{j}=\pi^{ik}\partial_{k}V^{j}-\bar{\Gamma}^{ij}_{k}V^{k},
\end{align}
where $V^{i}\partial_{i}$ is a vector field on the Poisson manifold.
A commutation relation of this covariant derivative is 
\begin{align}
[\bar{\nabla}^{i},\bar{\nabla}^{j}]V^{k}=-\bar{T}^{ij}_{m}\bar{\nabla}^{m}V^{k}+{\cal J}^{mij}\partial_{m}V^{k}-\bar{R}^{kij}_{m}V^{m}
\end{align}
where $\bar{T}$ and $\bar{R}$ are given by (\ref{eq:T}), (\ref{eq:R}), and
\begin{align}
{\cal J}^{mij}=&\pi^{mn}\partial_{n}\pi^{ij}+\pi^{in}\partial_{n}\pi^{jm}+\pi^{jn}\partial_{n}\pi^{mi}.
\end{align}
This tensor ${\cal J}^{ijk}$ is zero by the Poisson condition.
In the contravariant gravity, the torsionless condition is imposed:
\begin{align}
\bar{T}^{ij}_{k}
=0.
\end{align}
which determines the antisymmetric part of the connection.
We also impose a metric compatibility condition 
\begin{align}
\bar{\nabla}^{k}G^{ij}=0.
\end{align}
The unique connection in terms of the metric and the Poisson tensor is determined by solving the torsionless condition 
and the metric compatibility condition:
\begin{align}
	\bar{\Gamma}^{ij}_k
	&=\frac{1}{2}G_{mk}\left(\pi^{il} \p_l G^{jm}+\pi^{jl} \p_l G^{im}-\pi^{ml} \p_l G^{ij}
	+G^{lj}\p_l \pi^{mi}+G^{li}\p_l \pi^{mj}+G^{lm}\p_l \pi^{ij}\right).
\end{align}
The Ricci tensor and the Ricci scalar for the curvature tensor (\ref{eq:R}) are defined by
\begin{align}
\bar{R}^{ij}=\bar{R}^{ikj}_{k},\ \bar{R}=G_{ij}\bar{R}^{ij}.\label{eq:Riccis}
\end{align}
The contravariant Einstein equation is
\begin{align}
{\cal G}^{ij}=\bar{R}^{ij}-\frac{1}{2}G^{ij}\bar{R}=0.\label{eq:ceinsteineq}
\end{align}
Note that the tensor ${\cal G}^{ij}$ satisfies $G_{ij}\bar{\nabla}^{i}{\cal G}^{jk}=0$ because of the Bianchi identities of the contravariant Riemann tensor.

To formulate the equation of motion by the action principle, we assume a symplectic case 
in which the Poisson tensor is invertible.
We can construct an Einstein-Hilbert type theory:
\begin{align}
\int d^{n}x\sqrt{G^{-1}}e^{\phi}\bar{R}\label{jthiudnbgin}
\end{align}
where the integrand measure of this theory is determined by the divergence law:
\begin{align}
	e^\phi\sqrt{G^{-1}}  \bar{\nabla}^{i} \xi_i 
	&=\p_j (e^\phi\sqrt{G^{-1}}\xi_i \pi^{ij}  ),\label{eq:divl}
\end{align}
for any 1-form $\xi_{i}dx^{i}$.
This relation can be solved by
\begin{align}
e^{\phi}=\frac{1}{\det \pi \det G^{-1}}.
\end{align}
The action (\ref{jthiudnbgin}) gives the contravariant Einstein equation (\ref{eq:ceinsteineq}).

We can introduce a cosmological constant term to the Einstein-Hilbert type action.
\begin{align}
S_{c}=2\Lambda\int d^{n}e^{\phi}\sqrt{G^{-1}}.
\end{align}
The equation of motion with this cosmological constant term is
\begin{align}
\bar{R}^{ij}-\frac{1}{2}G^{ij}\bar{R}=\Lambda G^{ij}. \label{eomcc}
\end{align}
In the next section, we provide solutions of this equation of motion.

%%%%%%%%%%%%%%%%%%%%%%%%%%%%%%%%%%%%%%%%%%%%%%%
\subsection{Gauge symmetry}

The contravariant gravity, which is given by the Einstein-Hilbert type Lagrangian, is invariant under both 
diffeomorphism and $\beta$-diffeomorphism.

In the contravariant gravity we consider the Poisson tensor as a dynamical variable, 
and thus both the Poisson tensor and the metric are transformed as tensor under the diffeomorphism.
The Poisson-Lie derivative can also generate symmetry which is called $\beta$-diffeomorphism.
The gauge transformation of the $\beta$-diffeomorphism for the Poisson tensor and the metric is given by
\begin{align}
	\bar{\delta}_{\zeta} G^{ij}
	:=&\bar{\cal L}_{\zeta}G^{ij}\\
	=&\zeta_{k}\pi^{kl}\partial_{l}G^{ij}
	+(\pi^{ik}\partial_{k}\zeta_{l}+\zeta_{k}\partial_{l}\pi^{ik})G^{lj}
	+(\pi^{jk}\partial_{k}\zeta_{l}+\zeta_{k}\partial_{l}\pi^{jk})G^{il},\label{eq:bdiff1}\\
	\bar{\delta}_{\zeta} \pi^{ij}
	:=&0.\label{eq:bdiff2}
\end{align}
We note that the Poisson tensor is not a tensor $\bar{\delta}_{\zeta} \pi^{ij}\neq\bar{\cal L}_{\zeta}\pi$ under 
the $\beta$-diffeomorphism, because once we change the Poisson tensor underlining geometry totally change.

We can also understand this gauge transformation rule from physical point of view.
Here, let us write the Poisson-Lie derivative as $\bar{\cal L}^{\pi}_{\zeta}=\bar{\cal L}_{\zeta}$ 
to see the $\pi$ dependence explicitly.
If we assign $\bar{\delta}_{\zeta} \pi^{ij}=\bar{\cal L}^{\pi}_{\zeta}\pi^{ij}$ $i.e.$ the Poisson tensor is 
a tensor under the $\beta$-diffeomorphism, a commutation relation of the $\beta$-diffeomorphism transformation gives
\begin{align}
	[\bar{\delta}_{\eta},\bar{\delta}_{\zeta}]G^{ij}
	=&\bar{\delta}_{[\eta,\zeta]_{\pi}}G^{ij}+ {\cal L}^{\delta_{\eta}\pi}_{\zeta}G^{ij}-{\cal L}^{\delta_{\zeta}\pi}_{\eta}G^{ij}.
\end{align}
In this case, the algebra of this gauge symmetry fails to close.
Therefore, we should define (\ref{eq:bdiff2}) to realize the $\beta$-diffeomorphism in a field theory,
because a set of $\bar{\delta}$ and the fields is not a representation of the Poisson-Lie derivative.

Next, we consider the diffeomorphism in the contravariant gravity.
The diffeomorphism transformation rules for a coordinate transformation $x^{i}\rightarrow x'^{i}$ are given by
\begin{align}
	&G'^{ij}(x')=\frac{\partial x'^{i}}{\partial x^{m}} \frac{\partial x'^{j}}{\partial x^{n}} G^{mn}(x),\\
	&\pi'^{ij}(x')=\frac{\partial x'^{i}}{\partial x^{m}} \frac{\partial x'^{j}}{\partial x^{n}} \pi^{mn}(x).
\end{align}
For an infinitesimal diffeomorphism transformation $x'^{i}=x^{i}+X^{i}$, the transformation rules are
\begin{align}
	&\delta_{X} G^{ij}
	=X^{k}\partial_{k}G^{ij}-G^{il}\partial_{l}X^{j}-G^{jl}\partial_{l}X^{i},\\
	&\delta_{X} \pi^{ij}
	=X^{k}\partial_{k}\pi^{ij}-\pi^{il}\partial_{l}X^{j}+\pi^{jl}\partial_{l}X^{i}.
\end{align}
This gauge transformation is generated by the Lie derivative ${\cal L}_{X}$ with the vector field $X=X^{i}\partial_{i}$.
We stress that the Poisson tensor should transform together with the metric to get the correct transformation 
of the contravariant Levi-Civita connection:
\bea
	{(\bar{\Gamma}')}^{mn}_{l} 
	= \frac{\p x^k}{\p {x'}^{l}}
		\frac{\p {x'}^{m}}{\p x^i} \pi^{ij}\frac{\p^2 {x'}^{n}}{\p x^j \p x^k}
		+\frac{\p {x'}^{m}}{\p x^i}\frac{\p {x'}^{n}}{\p x^j}\frac{\p x^k}{\p {x'}^{l}}\bar{\Gamma}^{ij}_k.
		\label{trpG}
\ena
This formula can also be obtained from the axioms of the affine connection (\ref{affine}):
On another local patch, say $\{{x'}^i\}$, we have another set of basis 
of 1-forms $\{d{x'}^i\}$. For this basis, we introduce the coefficients $\bar{\Gamma}'$ as
\bea
	\bar{\nabla}_{d{x'}^{i}} d{x'}^{j}
			&\equiv {(\bar{\Gamma}')}^{ij}_{k} d{x'}^{k}.
\ena
On the intersection of the two local patches $\{{x}^i\}$ and $\{{x'}^i\}$,
we also have
\bea
	\bar{\nabla}_{d{x'}^{m}} d{x'}^{n}
			&= \bar{\nabla}_{\frac{\p {x'}^{m}}{\p x^i}dx^i} \bigg(\frac{\p {x'}^{n}}{\p x^j} dx^j\bigg)
			=\frac{\p{x'}^{m}}{\p x^i}\bigg(\ \pi^{ij}\frac{\p^2 {x'}^{n}}{\p x^j \p x^k}
				+\frac{\p {x'}^{n}}{\p x^j}\bar{\Gamma}^{ij}_k\bigg)dx^k,
\ena
by using \eqref{affine}  and \eqref{25}.
Then we can extract the behavior of the coefficients under coordinate transformations (\ref{trpG}).

Before closing this section, we mention a relation between the $\beta$-diffeomorphism $\bar{\delta}$ and the diffeomorphism $\delta$.
Since from the 1-form $\zeta$ and the Poisson tensor $\pi$ we can make a 1-vector $\bar{\iota}_{\zeta}\pi$, we see that
\begin{align}
&\bar{\delta}_{\zeta} G^{ij}-\delta_{\bar{\iota}_{\zeta}\pi}G^{ij}=\pi^{ik}(d\zeta)_{kl}G^{lj}+\pi^{jk}(d\zeta)_{kl}G^{li},\\
&\bar{\delta}_{\zeta} \pi^{ij}-\delta_{\bar{\iota}_{\zeta}\pi}\pi^{ij}=\pi^{ik}(d\zeta)_{kl}\pi^{lj}.
\end{align}
The right hand side can be understood as a gauge transformation of a B-field in string theory $B\rightarrow B+d\zeta$ 
after applying the open-closed relation with a strong B-field limit \cite{SeibergWitten,Blumenhagen:2012nt}.

%%%%%%%%%%%%%%%%%%%%%%%%%%%%%%%%%%%%%%%%%%%%
\subsection{Solutions of equation of motion}

We give some configurations which solve the equations of motion 
either with or without cosmological constant \eqref{eomcc}.
In this section we consider K\"ahler manifolds, $\mathbb{C}^{n}$, $\mathbb{C}P^{n}$, the Eguchi-Hanson space,
because these manifolds are implemented with a natural Poisson structure which is given by the inverse of the K\"ahler form.
For a K\"ahler potential $\phi$, the metric and the K\"ahler form are given by
\begin{align}
g_{\mu\bar{\nu}}
=
\frac{\partial^2\phi}{\partial z^{\mu}\partial \bar{z}^{\nu}},
~~~
\omega
=
ig_{\mu\bar{\nu}}dz^{\mu}\wedge d\bar{z}^{\nu},
\end{align}
where $\mu,\nu,...$ and $\bar{\mu},\bar{\nu},...$ denote indices of complex coordinates.
A canonical Poisson structure $\pi$ given 
by a matrix relation
\bea
	\pi^{ij}\omega_{jk}=\delta^i_k.
\ena

%%%%%%%%%%%%%%%%%%%%%%%%%%%%%%%%%%%%
\paragraph{Complex plane $\mathbb{C}^n$}

The simplest 
example is obtained by setting both the metric and Poisson tensor to be spatially constant.
In terms of the K\"ahler potential, 
the metric is specified by
\bea
	&\phi (z_1,\bar{z}_1,z_2,\bar{z}_2, \cdots,z_n,\bar{z}_n) =\frac{1}{2}(|z_1|^2+|z_2|^2+\cdots+|z_n|^2).
\ena
For the metric, we can obtain the corresponding K\"ahler form.
The inverse of the K\"ahler form gives the Poisson structure on $\mathbb{C}^n$.
As one easily sees that the contravariant Levi-Civita connection vanishes 
because it contains at least one derivative.
Therefore, the curvature tensor, Ricci tensor and the scalar curvature become trivial.
Thus this configuration is a solution to the equations of motion (\ref{eom1}).

%%%%%%%%%%%%%%%%%%%%%%%%%%%%%%%%%%%%
\paragraph{Complex Projective Space $\mathbb{C}P^n$}

A non-trivial solution is given by the complex projective space $\mathbb{C}P^n$.
We show explicit formulas for $\mathbb{C}P^1$ and $\mathbb{C}P^2$ and then we explain the generic $n$ case.

As is well known, the metric of the complex projective space $\mathbb{C}P^1$ 
is given by the Fubini-Study metric
\bea
			G=\frac{2dz\otimes d\bar{z}}{(|z|^2+1)^2},
\ena
and the symplectic form, i.e. the volume form for this case, is specified by
\bea
			\omega=\frac{2idz\wedge d\bar{z}}{(|z|^2+1)^2}.
\ena
In terms of K\"ahler potential the geometry is specified by
\bea
	&\phi (z,\bar{z}) =\log(1+|z|^2).
\ena
For these metric and the Poisson tensor we obtain
\bea
	 &\bar{R}^{{z}z\bar{z}}_{{z}} 
	= 2{(|z|^2+1)^2}, \\
	&\bar{R}^{z\bar{z}}
	=  2{(|z|^2+1)^2}, \\
	&\bar{R}=4.
\ena
Then we see that the equation of motion without cosmological constant \eqref{eom1} is satisfied
\bea
	\bar{R}^{z\bar{z}} - \frac{1}{2}G^{z\bar{z}}\bar{R} =2{(|z|^2+1)^2}
	- \frac{1}{2} \cdot(|z|^2+1)^2 \cdot 4 =0.
\ena
It is also an example of the contravariant K\"ahler-Einstein manifold:
\bea
	\bar{R}^{z\bar{z}} = 2G^{z\bar{z}}.
\ena

For the two-dimensional complex projective space $\mathbb{C}P^2$ the corresponding K\"ahler potential is
\bea
	&\phi (z_1,\bar{z}_1,z_2,\bar{z}_2) =\log(1+|z_1|^2+|z_2|^2).
\ena
The metric of the complex projective space $\mathbb{C}P^2$ 
is given by
\bea
		G_{ij}=\frac{1}{(|z_1|^2+|z_2|^2+1)^2}\left(%
		\begin{array}{cccc} 
		0 & |z_2|^2+1  &  0  &  -\bar{z}_1z_2 \\
		|z_2|^2+1  &  0  &   -z_1\bar{z}_2 &   0 \\
		0  & -z_1\bar{z}_2  & 0  &  |z_1|^2+1  \\
		 -\bar{z}_1z_2  &  0 &  |z_1|^2+1  & 0 
		 \end{array}\right),%,
\ena
where the matrix entry $i=1,2,3,4$ corresponds to $z_1, \bar{z}_1, z_2, \bar{z}_2$, respectively.
And the K\"ahler form,  a symplectic form associated with a K\"ahler manifold, is 
\bea
			\omega_{ij}=\frac{i}{(|z_1|^2+|z_2|^2+1)^2}\left(%
		\begin{array}{cccc} 
		0 & |z_2|^2+1  &  0  &  -\bar{z}_1{z}_2 \\
		-(|z_2|^2+1)  &  0  &   {z}_1\bar{z}_2 &   0 \\
		0  & - {z}_1\bar{z}_2 & 0  &  |z_1|^2+1  \\
		 \bar{z}_1{z}_2  &  0 &  -(|z_1|^2+1 ) & 0 
		 \end{array}\right),%,
\ena
and its corresponding canonical Poisson structure $\pi$ is
\bea
	\pi^{ij}=( |z_1|^2+|z_2|^2+1)\left(%
		\begin{array}{cccc} 
		0 & i (|z_1|^2+1)  &  0  &  i{z}_1\bar{z}_2 \\
		-i (|z_1|^2+1)  &  0  &   -i\bar{z}_1{z}_2 &   0 \\
		0  & i \bar{z}_1{z}_2 & 0  &  i (|z_2|^2+1)   \\
		 -i{z}_1\bar{z}_2  &  0 &  -i (|z_2|^2+1) & 0 
		 \end{array}\right).%,
\ena
Then after some computation, we see that it turns out to be an Einstein manifold:
\bea
	\bar{R}^{ij}=3G^{ij}.
\ena
Hence this configuration is a solution to the equation of motion \eqref{eom1} with a cosmological constant
$\Lambda = -3$.

Finally, we show that the complex projective space $\mathbb{C}P^n$ is a solution of the contravariant Einstein equation.
The K\"ahler potential of this space is given by
\begin{align}
\phi (z_1,\bar{z}_1,z_2,\bar{z}_2, \cdots,z_n,\bar{z}_n) =\log\left(1+\sum_{m=1}^{n}|z_{m}|^2\right).
\end{align}
We can easily show that 
\begin{align}
	\bar{R}^{ij}=(n+1)G^{ij},
\end{align}
and this space solve the contravariant Einstein equation with a cosmological constant $\Lambda=1-n^2$.

%%%%%%%%%%%%%%%%%%%%%%%%%%%%%%%%%%%%%%%
\paragraph{Eguchi-Hanson Space}

Another example is that known as Eguchi-Hanson space, which is Ricci flat.
This solution is originally considered
in the context of gravitational instanton.
Its K\"ahler potential is given by
\bea
	\phi (z_1,\bar{z}_1,z_2,\bar{z}_2)
	=\log \bigg(
	\frac{\rho^2e^{\sqrt{\rho^4+a^4}/a^2}}{a^2+\sqrt{\rho^4+a^4}} \bigg),
\ena
with $a$ is a real constant and 
\bea
	\rho^2 =|z_1|^2+|z_2|^2.
\ena
We do not put the explicit forms of the metric nor the K\"ahler form here
and omit to present the detailed computational process,
but we can actually show that its contravariant Ricci tensor has trivial components
\bea
	\bar{R}^{ij}=0.
\ena

%%%%%%%%%%%%%%%%%%%%%%%%%%%%%%%%%%%%%%%
\subsection{Linearization}

In this subsection, we consider fluctuations of the metric and the Poisson tensor around a 
background configuration $(G^{ij},\pi^{ij})$, which solves the Einstein equation \eqref{eomcc}.
We assume that the fluctuation of the Poisson tensor dose not break the Poisson condition.
In \cite{Hawkins:2002rf}, the metric fluctuation on the fixed Poisson background is discussed.
It is important to introduce the Poisson tensor fluctuation to discuss symmetry and its relation 
with the noncommutative gauge theory in the next section.

The fluctuation of the metric and the Poisson tensor are
\begin{align}
G^{ij}+\epsilon h^{ij},~~\pi^{ij}+\epsilon\rho^{ij},
\end{align}
where $\epsilon$ is an infinitesimal constant.
We consider the first order approximation of the contravariant Riemann curvature at first.
The contravariant Levi-Civita connection can be approximated by
\begin{align}
\bar{\Gamma}^{ij}_{k}(G^{ij}+\epsilon h^{ij},\pi^{ij}+\epsilon\rho^{ij})
=&	\bar{\Gamma}^{ij}_{k}(G^{ij},\pi^{ij})\nonumber\\
&	-\epsilon\rho^{im}\Gamma^{j}_{mk}\nonumber\\
&	+\frac{\epsilon}{2}G_{km}
		(\bar{\nabla}^{i}h^{jm}
		+\bar{\nabla}^{j}h^{im}
		-\bar{\nabla}^{m}h^{ij})\nonumber\\
&	+\frac{\epsilon}{2}G_{km}
		(G^{in}\nabla_{n}\rho^{mj}
		+G^{jn}\nabla_{n}\rho^{mi}
		+G^{mn}\nabla_{n}\rho^{ij})+{\cal O}(\epsilon^2),
\end{align}
where $\Gamma^{i}_{jk}$ is an ordinary Levi-Civita connection which is given by $G^{-1}$ 
and $\nabla$ is a covariant derivative with the connection $\Gamma^{i}_{jk}$.
Using this connection we obtain the linearized contravariant Riemann tensor.
\begin{align}
\bar{R}^{kij}_{l}(G^{ij}+\epsilon h^{ij},\pi^{ij}+\epsilon\rho^{ij})
=&	\bar{R}^{kij}_{l}(G^{ij},\pi^{ij})\nonumber\\
&	+\frac{\epsilon}{2}G_{ln}
		(\bar{\nabla}^{i}\bar{\nabla}^{k}h^{jn}
		-\bar{\nabla}^{i}\bar{\nabla}^{n}h^{jk}
		-\bar{\nabla}^{j}\bar{\nabla}^{k}h^{in}
		+\bar{\nabla}^{j}\bar{\nabla}^{n}h^{ik})\nonumber\\
&	-\frac{\epsilon}{2}G_{ln}
		(\bar{R}^{kij}_{m}h^{mn}
		+\bar{R}^{nij}_{m}h^{km})\nonumber\\
&	+\frac{\epsilon}{2}G_{ln}
		(G^{jm}\bar{\nabla}^{i}\nabla_{m}\rho^{nk}
		+G^{km}\bar{\nabla}^{i}\nabla_{m}\rho^{nj}
		+G^{mn}\bar{\nabla}^{i}\nabla_{m}\rho^{jk})\nonumber\\
&	-\frac{\epsilon}{2}G_{ln}
		(G^{im}\bar{\nabla}^{j}\nabla_{m}\rho^{nk}
		+G^{km}\bar{\nabla}^{j}\nabla_{m}\rho^{ni}
		+G^{mn}\bar{\nabla}^{j}\nabla_{m}\rho^{ik})\nonumber\\
&	-\epsilon\rho^{im}\pi^{jn}R^{k}_{lmn}
	+\epsilon\rho^{jm}\pi^{in}R^{k}_{lmn}\nonumber\\
&	+\epsilon\rho^{im}\nabla_{m}K^{jk}_{l}
	-\epsilon\rho^{jm}\nabla_{m}K^{ik}_{l}
	-\epsilon\nabla_{m}\rho^{ij}K^{mk}_{l}+{\cal O}(\epsilon^2),\label{eq:fulllinear}
\end{align}
where the contorsion tensor $K^{ij}_{k}$ is defined by
\begin{align}
K^{ij}_{k}=\frac{1}{2}G_{kl}(\nabla^{l}\pi^{ij}-\nabla^{i}\pi^{jl}+\nabla^{j}\pi^{li}).
\end{align}
In this formula, we used the Poisson condition for $\pi+\epsilon\rho$.
See appendix \ref{sec:linearapp} for detail calculations.

So far we have considered fluctuations around a general background.
In the following part of this section, we discuss fluctuation around a flat metric $\delta^{ij}$, $G^{ij}=\delta^{ij}+\epsilon h^{ij}$, 
and a constant Poisson background $\theta^{ij}$, $\pi^{ij}=\theta^{ij}+\epsilon \rho^{ij}$.
Around the flat background, the contravariant Levi-Civita is
\begin{align}
	\bar{\Gamma}^{ij}_{k}(\delta^{ij}+\epsilon h^{ij},\theta^{ij}+\epsilon \rho^{ij})
	=&\frac{1}{2}\epsilon(
		\theta^{il}\partial_{l}h^{j}{}_{k}+\theta^{jl}\partial_{l}h^{i}{}_{k}-\theta_{k}{}^{l}\partial_{l}h^{ij}
		+\partial^{j}\rho_{k}{}^{i}+\partial^{i}\rho_{k}{}^{j}+\partial_{k}\rho^{ij})	\label{eq:linearLC}
\end{align}
where indices are raised and lowered by $\delta^{ij}$ not $\theta^{ij}$.
Using (\ref{eq:fulllinear}), we get
\begin{align}
&\bar{R}^{kij}_{l}(\delta^{ij}+\epsilon h^{ij},\theta^{ij}+\epsilon \rho^{ij})\nonumber\\
=&\frac{\epsilon}{2}
(
\theta^{kn}\theta^{im}\partial_{n}\partial_{m}h^{j}{}_{l}-\theta_{l}{}^{n}\theta^{im}\partial_{n}\partial_{m}h^{jk}
+\theta^{im}\partial^{k}\partial_{m}\rho_{l}{}^{j}+\theta^{im}\partial^{j}\partial_{m}\rho_{l}{}^{k}+\theta^{im}\partial_{l}\partial_{m}\rho^{jk}
)\nonumber\\
&-\frac{\epsilon}{2}
(
\theta^{kn}\theta^{jm}\partial_{n}\partial_{m}h^{i}{}_{l}-\theta_{l}{}^{n}\theta^{jm}\partial_{n}\partial_{m}h^{ik}
+\theta^{jm}\partial^{k}\partial_{m}\rho_{l}{}^{i}+\theta^{jm}\partial^{i}\partial_{m}\rho_{l}{}^{k}+\theta^{jm}\partial_{l}\partial_{m}\rho^{ik}
)+{\cal O}(\epsilon^2).
\end{align}
The contravariant Ricci tensor can be written by
\begin{align}
	\bar{R}^{kj}(\delta^{ij}+\epsilon h^{ij},\theta^{ij}+\epsilon \rho^{ij})
	=&\frac{\epsilon}{2}
		(\theta^{kn}\theta^{im}\partial_{n}\partial_{m}h^{j}{}_{i}
		+\theta^{jn}\theta^{im}\partial_{n}\partial_{m}h^{k}{}_{i}
		-\theta_{i}{}^{n}\theta^{im}\partial_{n}\partial_{m}h^{jk}
		-\theta^{kn}\theta^{jm}\partial_{n}\partial_{m}h^{i}{}_{i})	\nonumber\\
	&+\frac{\epsilon}{2}
		(\theta^{im}\partial^{k}\partial_{m}\rho_{i}{}^{j}
		+\theta^{im}\partial^{j}\partial_{m}\rho_{i}{}^{k}
		-\theta^{jm}\partial_{i}\partial_{m}\rho^{ik}
		-\theta^{km}\partial_{i}\partial_{m}\rho^{ij})+{\cal O}(\epsilon^2).
\end{align}
Finally, the contravariant Ricci scalar is
\begin{align}
	\bar{R}(\delta^{ij}+\epsilon h^{ij},\theta^{ij}+\epsilon \rho^{ij})
		=&\epsilon
			(\theta^{jn}\theta^{im}\partial_{n}\partial_{m}h_{ij}
			-\theta_{j}{}^{n}\theta^{jm}\partial_{n}\partial_{m}h^{i}{}_{i}
			+2\theta^{im}\partial_{j}\partial_{m}\rho_{i}{}^{j})+{\cal O}(\epsilon^2),
\end{align}
Introducing $ \tilde{\partial}^{i}=\theta^{ij}\partial_{j}$,
the Ricci tensor and the Ricci scalar are
\begin{align}
\bar{R}^{kj}(\delta^{ij}+\epsilon h^{ij},\theta^{ij}+\epsilon \rho^{ij})=
	&\frac{\epsilon}{2}
		(\tilde{\partial}^{k}\tilde{\partial}^{i}h^{j}{}_{i}+\tilde{\partial}^{j}\tilde{\partial}^{i}h^{k}{}_{i}
		-\tilde{\partial}_{i}\tilde{\partial}^{i}h^{jk}
		-\tilde{\partial}^{j}\tilde{\partial}^{k}h^{i}{}_{i})	\nonumber\\
	&+\frac{\epsilon}{2}
		(\partial^{k}\tilde{\partial}^{i}\rho_{i}{}^{j}+\partial^{j}\tilde{\partial}^{i}\rho_{i}{}^{k}
		-\partial_{i}\tilde{\partial}^{j}\rho^{ik}-\partial_{i}\tilde{\partial}^{k}\rho^{ij})+{\cal O}(\epsilon^2),	\\
\bar{R}(\delta^{ij}+\epsilon h^{ij},\theta^{ij}+\epsilon \rho^{ij})=
	&\epsilon
		(\tilde{\partial}^{i}\tilde{\partial}^{j}h_{ij}
		-\tilde{\partial}_{j}\tilde{\partial}^{j}h^{i}{}_{i}
		-2\partial_{i}\tilde{\partial}^{j}\rho^{i}{}_{j})+{\cal O}(\epsilon^2).
\end{align}
Therefore, the contravariant Einstein equation is simplified in the following form
\begin{align}
	(\tilde{\partial}^{k}\tilde{\partial}_{i}h^{ji}+\tilde{\partial}^{j}\tilde{\partial}_{i}h^{ki}
	-\tilde{\partial}^{i}\tilde{\partial}_{i}h^{jk}
	-\tilde{\partial}^{j}\tilde{\partial}^{k}h_{i}{}^{i})
	-\delta^{kj}
		(\tilde{\partial}_{i}\tilde{\partial}_{l}h^{il}-\tilde{\partial}^{l}\tilde{\partial}_{l}h_{i}{}^{i})\nonumber\\
	+(\partial^{k}\tilde{\partial}^{i}\rho_{i}{}^{j}+\partial^{j}\tilde{\partial}^{i}\rho_{i}{}^{k}
	-\partial_{i}\tilde{\partial}^{j}\rho^{ik}-\partial_{i}\tilde{\partial}^{k}\rho^{ij})
	+2\delta^{kj}\partial_{i}\tilde{\partial}^{j}\rho^{i}{}_{j}
	=0.\label{eq:ssss}
\end{align}

As the usual linearized Einstein equation, we can find that
\begin{align}
\tilde{\partial}_{j}((\tilde{\partial}^{k}\tilde{\partial}_{i}h^{ji}+\tilde{\partial}^{j}\tilde{\partial}_{i}h^{ki}
	-\tilde{\partial}^{i}\tilde{\partial}_{i}h^{jk}
	-\tilde{\partial}^{j}\tilde{\partial}^{k}h_{i}{}^{i})
	-\delta^{kj}
		(\tilde{\partial}_{i}\tilde{\partial}_{l}h^{il}-\tilde{\partial}^{l}\tilde{\partial}_{l}h_{i}{}^{i}))=0.
\end{align}
This equation implies that the remaining part of the equation (\ref{eq:ssss}) should be divergenceless.
\begin{align}
0=&\tilde{\partial}_{j}((\partial^{k}\tilde{\partial}^{i}\rho_{i}{}^{j}+\partial^{j}\tilde{\partial}^{i}\rho_{i}{}^{k}
	-\partial_{i}\tilde{\partial}^{j}\rho^{ik}-\partial_{i}\tilde{\partial}^{k}\rho^{ij})
	+2\delta^{kj}\partial_{i}\tilde{\partial}^{j}\rho^{i}{}_{j})\nonumber\\
=&\tilde{\partial}_{j}\partial_{i}(-\tilde{\partial}^{j}\rho^{ik}+\tilde{\partial}^{k}\rho^{ij}).\label{eq:poissonltt}
\end{align}
Since this relation is satisfied by a linearized version of the Poisson condition:
\begin{align}
\tilde{\partial}^{i}\rho^{jk}+\tilde{\partial}^{j}\rho^{ki}+\tilde{\partial}^{k}\rho^{ij}=0\label{eq:lPoisson}
\end{align}
this provides us with a consistency check for the linearized contravariant Einstein equation (\ref{eq:ssss}) 
with the fluctuation of the Poisson tensor $\rho$.

Before closing this section, we discuss a linearized version of the gauge symmetries of the contravariant gravity.
In terms of $h^{ij}$ and $\rho^{ij}$, the diffeomorphism transformations are
\begin{align}
	&\delta_{X}h^{ij}=-\partial^{i}X^{j}-\partial^{j}X^{i},\label{eq:ponta1}\\
	&\delta_{X}\rho^{ij}=-\tilde{\partial}^{i}X^{j}+\tilde{\partial}^{j}X^{i},\label{eq:ponta2}
\end{align}
on the other hand, the $\beta$-diffeomorphism transformations are
\begin{align}
	&\bar{\delta}_{\zeta}h^{ij}=\tilde{\partial}^{i}\zeta^{j}+\tilde{\partial}^{j}\zeta^{i},\label{eq:bnanaco}\\
	&\bar{\delta}_{\zeta}\rho^{ij}=0.
\end{align}
It is interesting that the symmetry transformation (\ref{eq:bnanaco}) looks like a diffeomorphism of 
the linearized Einstein equation (\ref{eq:ssss}),
on the other hand (\ref{eq:ponta1}) and (\ref{eq:ponta2}) are diffeomorphism in terms of original symmetry of the contravariant gravity theory.

We comment on the fluctuation of the Poisson tensor.
In general, the deformation of the Poisson structure is a very difficult problem comparing to the deformation of the symplectic structure\cite{Vaisman1,BouisaghouaneKiselev,BouisaghouaneKiselevRutten,Kontsevich}.
Here, our consideration is restricted on the linearization level.
On the other hand, in this paper we have not given any discussion on the causal structure of this theory.
Since the derivations in the contravariant geometry are contracted with the Poisson tensor, 
we will have to reconsider the causal structure which is suitable for this geometry and even for 
the noncommutative geometry \cite{Hawkins:2002rf}.

%%%%%%%%%%%%%%%%%%%%%%%%%%%%%%%%%%%%%%%%%%%%
\section{Contravariant geometry and noncommutative gauge theory on Moyal Plane}

In this section, we discuss a relation between the contravariant geometry and the emergent gravity 
from the noncommutative gauge theory on the Moyal Plane.
First, we briefly review the idea of the emergent gravity which is based on
the relation between gravity and the noncommutative U(1) gauge theory via the Seiberg-Witten map.
Then, we show that the contravariant geometry is an appropriate framework for describing the emergent gravity 
by constructing the contravariant Ricci tensor for the resulting metric given in the emergent gravity scenario.

%%%%%%%%%%%%%%%%%%%%%%%%%%%%%%%%%%%%%%%%%%%%
\subsection{Noncommutative gauge theory on Moyal plane}

Let us consider a $D$-dimensional noncommutative space where the star product $f*g$ 
between functions $f,g$ is defined, and the star commutator is denoted as
\begin{align}
[f,g]_{*}=f*g-g*f.
\end{align}
In terms of the local coordinate $x^{i}$, the noncommutativity is characterized by
\begin{align}
[x^{i},x^{j}]_{*}=i\theta^{ij}.\label{eq:xxt}
\end{align}
Here, we assume that the antisymmetric tensor $\theta^{ij}$ does not depend on $x^{i}$.
This relation is realized by the Moyal star product
\begin{align}
(f*g)(x)
=&f(x)\exp\left(\frac{i}{2}\overleftarrow{\partial_{i}}\theta^{ij}\overrightarrow{\partial_{j}}\right)g(x).
\end{align}

We consider the gauge theory on the Moyal plane.
For the noncommutative gauge field $\hat{A}_{i}$, the gauge transformation with a gauge parameter $\hat{\lambda}$ is
\begin{align}
\hat{\delta}_{\hat{\lambda}} \hat{A}_{i}=\partial_{i}\hat{\lambda}-i[\hat{A}_{i},\hat{\lambda}]_{*}.
\end{align}
The field strength for the gauge field is given by
\begin{align}
\hat{F}_{ij}=\partial_{i}\hat{A}_{j}-\partial_{j}\hat{A}_{i}-i[\hat{A}_{i},\hat{A}_{j}]_{*}.
\end{align}
This field strength is covariant under the gauge transformation $\delta \hat{F}_{ij}=-i[\hat{F}_{ij},\hat{\lambda}]_{*}$.
The gauge invariant kinetic term of the U(1) gauge field is given by
\begin{align}
S_{\hat{A}}=\frac{1}{4}\int d^{D}x\hat{F}_{ij}*\hat{F}^{ij},
\end{align}
where indices are contracted by the flat metric $\delta_{ij}$.

%%%%%%%%%%%%%%%%%%%%%%%%%%%%%%%%%%%%%%%%%%%%
\subsection{Seiberg-Witten map and emergent metric}

The Seiberg-Witten map tells us existence of a relation between the noncommutative gauge field $\hat{A}_{i}$ 
and the ordinary gauge field $A_{i}$ as \cite{SeibergWitten}
\begin{align}
	&\hat{A}=\hat{A}(A),\
	\hat{\lambda}=\hat{\lambda}(\lambda,A).
\end{align}
The Seiberg-Witten map is characterized by a condition
\begin{align}
	&\hat{A}_{i}(A)+\delta_{\hat{\lambda}}\hat{A}_{i}(A)=\hat{A}_{i}(A+\delta_{\lambda}A).\label{enbbnsfbjlxg}
\end{align}
where $\delta_{\lambda} A_{i}=\partial_{i}\lambda$.
The condition (\ref{enbbnsfbjlxg}) can be solved order by order of the deformation parameter $\theta^{ij}$.
An explicit expression of the Seiberg-Witten map of the first order of $\theta^{ij}$ is given by
\begin{align}
&\hat{A}_{i}
=
A_{i}-\frac{1}{2}\theta^{kl}A_{k}(\partial_{l}A_{i}+F_{li})+{\cal O}(\theta^{2}),\label{eq:hatA}\\
&\hat{\lambda}
=
\lambda-\frac{1}{2}\theta^{kl}A_{k}\partial_{l}\lambda+{\cal O}(\theta^{2}),
\end{align}
where $F_{ij}=\partial_{i}A_{j}-\partial_{j}A_{i}$.
By using (\ref{eq:hatA}), the noncommutative field strength can be calculated as
\begin{align}
\hat{F}_{ij}=
F_{ij}
-\theta^{kl}F_{ik}F_{lj}-\theta^{kl}A_{k}\partial_{l}F_{ij}+{\cal O}(\theta^{2}).
\end{align}
For this field strength the kinetic term of the noncommutative gauge theory can be written by 
\begin{align}
S_{\hat{A}}
=&\frac{1}{4}\int d^{D}x\left(F_{ij}F^{ij}+2\theta^{kl}F_{l}{}^{m}\left(F_{m}{}^{i}F_{ik}+\frac{1}{4}\delta_{km}F^{ij}F_{ij}\right)+{\cal O}(\theta^{2})\right).\label{eq:pizza}
\end{align}

It is known that this action of the noncommutative U(1) gauge theory (\ref{eq:pizza}) can be rewritten as \cite{Rivelles}
\begin{align}
S_{g,A}=\frac{1}{4}\int d^{D}x\sqrt{g}g^{ij}g^{kl}F_{ik}F_{jl},\label{ndvxjfbvxn}
\end{align}
with
\begin{align}
&g_{ij}
=
\delta_{ij}+u_{ij},\\
&u^{ij}
=
\frac{1}{2}\theta^{ik}F_{k}{}^{j}
+\frac{1}{2}\theta^{jk}F_{k}{}^{i}
+\frac{\kappa_{0}}{4}\delta^{ij}\theta^{kl}F_{kl},\label{eq:h}
\end{align}
where $\kappa_{0}$ is a constant which depends on $D$.
This action (\ref{ndvxjfbvxn}) is interpreted as a kinetic term of the U(1) gauge theory on a curved background with a metric $g_{ij}$ \cite{Rivelles}, see also \cite{Yang,Steinacker,Steinacker3,YS,Y}.
We consider the U(1) gauge field in this paper and we take\footnote{In the literature \cite{Rivelles} the constant $\kappa_{0}$ is chosen as $\kappa_{0}=1$ in $D=4$. When we consider the $D=4$ case, there is an ambiguity of a choice of $\kappa_{0}$ in (\ref{eq:h}). 
Since the U(1) gauge theory (\ref{eq:pizza}) is Weyl invariant under the local scaling of the flat metric, $\kappa_{0}$ in (\ref{eq:h}) of the resulting metric can be changed by the Weyl transformation $g_{ij}\rightarrow e^{\phi}g_{ij}$.
And thus $\kappa_{0}$ can be taken zero.
Except for $D=4$,  we should choose $\kappa_{0}=0$.}
\begin{align}
\kappa_{0}=0.
\end{align}
The standard Ricci tensor and scalar of the resulting metric (\ref{eq:h}) are \cite{Rivelles}
\begin{align}
&R_{ij}(\delta_{ij}+u_{ij})
=
-\frac{1}{2}(
	\partial^{2}\tilde{\partial}_{i}A_{j}
	+\partial^{2}\tilde{\partial}_{j}A_{i}
	+\partial_{i}\tilde{\partial}_{j}\partial_{k}A^{k}
	+\partial_{j}\tilde{\partial}_{i}\partial_{k}A^{k})
	-\partial_{i}\partial_{j}\tilde{\partial}^{k}A_{k}
+{\cal O}(\theta^2),\label{eq:RicciNC}\\
&R(\delta_{ij}+u_{ij})
=
-2\partial^{2}\tilde{\partial}^{k}A_{k}
+{\cal O}(\theta^2),
\end{align}
where $\tilde{\partial}^{i}=\theta^{ij}\partial_{j}$.

One can consider introducing other fields, for instance a scalar field.
In 4-dimensional case, when one adds the scalar field kinetic term to the action (\ref{eq:pizza}), it breaks
invariance under the local scaling of the flat metric.
This scaling rule determines the nonzero value of $\kappa_{0}$ as discussed in \cite{Rivelles}.
The noncommutative gauge theories given by matrix models also allow gravitational interpretation \cite{Steinacker}.

%%%%%%%%%%%%%%%%%%%%%%%%%%%%%%%%%%%%%%%%%%%%
\subsection{Relation with contravariant geometry}

In this section, we discuss relations with contravariant gravity and noncommutative gauge theory.
A role of the fluctuation of the Poisson tensor in the linearized contravariant Ricci tensor becomes clear 
in the scenario of the emergent gravity.

In the linear approximation of the contravariant geometry, the Poisson condition is given by (\ref{eq:lPoisson}).
Thus, the linearized Poisson condition can be interpreted as the Bianchi identity of the gauge theory.
This suggests that the fluctuation $\rho$ is identified with the gauge field $A^{i}$ in the following way
\begin{align}
\rho^{ij}=\tilde{\partial}^{i}A^{j}-\tilde{\partial}^{j}A^{i}.\label{eq:frafra}
\end{align}
Here, we consider the metric as a background, we take the metric fluctuation in the contravariant gravity as
\begin{align}
h^{ij}=0.
\end{align}
By using the relation (\ref{eq:frafra}), the contravariant Ricci tensor is
\begin{align}
\bar{R}^{jk}(\delta^{ij},\theta^{ij}+\tilde{\partial}^{i}A^{j}-\tilde{\partial}^{j}A^{i})
=&\tilde{\partial}^2\partial^{j}A^{k}+\tilde{\partial}^2\partial^{k}A^{j}-\partial^{j}\tilde{\partial}^{k}\tilde{\partial}^{i}A_{i}-\partial^{k}\tilde{\partial}^{j}\tilde{\partial}^{i}A_{i}+2\tilde{\partial}^{j}\tilde{\partial}^{k}\partial_{i}A^{i}.
\end{align}
We can compare the result of the noncommutative gauge theory (\ref{eq:RicciNC}).
\begin{align}
-\bar{R}_{ij}(\delta^{ij},\theta^{ij}+\tilde{\partial}^{i}A^{j}-\tilde{\partial}^{j}A^{i})=R_{ij}(\delta^{ij}+\theta^{ik}F_{k}{}^{j}+\theta^{jk}F_{k}{}^{i})|_{\partial_{i}\leftrightarrow \tilde{\partial}_{i}}
\label{eq:curvaturecg}
\end{align}
where the left hand side is calculated from the contravariant gravity, and the right hand side is obtained by the noncommutative gauge theory.
Since the metric fluctuation is treated as a background, the nontrivial curvature can be recovered by the Poisson tensor.
The equation (\ref{eq:curvaturecg}) shows that we can identify the fluctuation of the Poisson tensor $\rho$ in the contravariant geometry with the fluctuation of the emergent metric $u$
by interchanging the role of derivative $\partial$ and $\tilde{\partial}$ in this expression.
As we will see in the K\"ahler manifold case this identification is more natural.
For completeness we also discuss the relation between the contravariant gravity and the emergent gravity obtained from the matrix model in Appendix C.
Therefore, the contravariant geometry
is an appropriate geometric framework for the emergent gravity from the noncommutative gauge theory.

%%%%%%%%%%%%%%%%%%%%%%%%%%%%%%%%%%%%%%%%%%%%
\section{Contravariant geometry and noncommutative gauge theory on homogeneous K\"ahler manifold}

In this section, we argue the relation between the contravariant geometry and 
the emergent geometry which we obtained from the gauge theory on the noncommutative K\"ahler manifold.
First, we review a construction of noncommutative gauge theories on a  homogeneous K\"ahler manifold.
Then, we derive an emergent metric starting from the noncommutative theory.
We show that the curvature of the contravariant geometry coincides with the one from the noncommutative gauge theory.

%%%%%%%%%%%%%%%%%%%%%%%%%%%%%%%%%%%%%%%%%%%%
\subsection{Noncommutative gauge theory on K\"ahler manifold}

Our geometrical setup is the following:
Let $\Phi$ be a K\"ahler potential of a homogeneous K\"ahler manifold $({\cal M},g)$, ${\cal M}\simeq G/H$.
The metric $g$ and the K\"ahler form $\omega$ can be written by
\begin{align}
g_{\mu\bar{\nu}}
=
\frac{\partial^2\Phi}{\partial z^{\mu}\partial \bar{z}^{\nu}},
~~~
\omega
=
ig_{\mu\bar{\nu}}dz^{\mu}\wedge d\bar{z}^{\nu}.
\end{align}
The index $\mu$ 
 runs from $1$ to $N$.
We also use $i=\{\mu,\bar{\mu}\}$ which runs $1$ to $2N$.
In this case the K\"ahler manifold has Killing vectors ${\cal L}_{a}=\zeta^{i}_{a}\partial_{i}=\zeta^{\mu}_{a}\partial_{\mu}+\zeta^{\bar{\mu}}_{a}\partial_{\bar{\mu}}$, $a,b,...=1,...\text{dim}G$, which satisfies
\begin{align}
[{\cal L}_{a},{\cal L}_{b}]=if_{ab}^{c}{\cal L}_{c},
\end{align}
where $f^{c}_{ab}$ is a structure constant of the isometry group of the K\"ahler manifold.
We also assume these Killing vectors are (anti-)holomorphic $\partial_{\mu}\zeta^{\bar{\nu}}_{a}=\partial_{\bar{\mu}}\zeta^{\nu}_{a}=0$.
For any 1-form $\alpha=\alpha_{\mu}dz^{\mu}+\alpha_{\bar{\mu}}d\bar{z}^{\mu}$, we define $\alpha_{a}=\iota_{{\cal L}_{a}}\alpha=\zeta^{i}_{a}\alpha_{i}=\zeta^{\mu}_{a}\alpha_{\mu}+\zeta^{\bar{\mu}}_{a}\alpha_{\bar{\mu}}$.
The isometry group has a Killing form $g^{ab}$ which satisfies $g^{ab}\zeta^{\mu}_{b}\zeta^{\bar{\nu}}_{a}=g^{\bar{\nu}\mu}$.
Explicit examples are given in \cite{Sako}.

The noncommutative U(1) gauge theory on the homogeneous K\"ahler manifold is constructed in \cite{MSSU}.
The kinetic term for the gauge field is given by\footnote{The measure of the action is determined by the trace property which depends on a detail of the star product on the background manifold.}
\begin{align}
S=\frac{1}{4}\int d^{N}zd^{N}\bar{z}\sqrt{g}g^{ab}g^{cd}\hat{F}_{ac}*\hat{F}_{bd}.
\end{align}
where the field strength for the gauge field $\hat{A}$ is defined by
\begin{align}
\hat{F}_{ab}={\cal L}_{a}\hat{A}_{b}-{\cal L}_{b}\hat{A}_{a}-\hat{A}_{a}*\hat{A}_{b}+\hat{A}_{b}*\hat{A}_{a}-if_{ab}^{c}\hat{A}_{c}.
\end{align}
The leading order of noncommutative product $*$ for functions $f$ and $g$ is given by
\begin{align}
f*g
=&
fg+\hbar g^{\bar{\mu}\nu}\partial_{\bar{\mu}}f\partial_{\nu}g+{\cal O}(\hbar^2),
\end{align}
where we use the noncommutative parameter $\hbar$ instead of $\theta$.
For this star product we can show that
\begin{align}
{\cal L}_{a}(f*g)=({\cal L}_{a}f)*g+f*{\cal L}_{a}g,
\end{align}
by using the Killing equation
\begin{align}
{\cal L}_{{\cal L}_{a}}g^{\rho\bar{\sigma}}=\zeta^{\mu}_{a}\partial_{\mu}g^{\rho\bar{\sigma}}+\zeta^{\bar{\mu}}_{a}\partial_{\bar{\mu}}g^{\rho\bar{\sigma}}-g^{\mu\bar{\sigma}}\partial_{\mu}\zeta^{\rho}_{a}-g^{\rho\bar{\mu}}\partial_{\bar{\mu}}\zeta^{\bar{\sigma}}_{a}=0.
\end{align}
The gauge transformation is given by
\begin{align}
\hat{A}_{a}\rightarrow \hat{A}'_{a}=iU^{-1}*{\cal L}_{a}U+U^{-1}*\hat{A}_{a}*U.
\end{align}
The field strength is covariant under the gauge transformation $\hat{F}'_{ab}=U^{-1}*\hat{F}_{ab}*U$.
Under the infinitesimal transformation $U=1-i\hat{\lambda}$, the gauge transformation is
\begin{align}
\delta \hat{A}_{a}={\cal L}_{a}\hat{\lambda}-i\hat{A}_{a}*\hat{\lambda}+i\hat{\lambda}*\hat{A}_{a}.
\end{align}
The action is invariant under this gauge transformation.

%%%%%%%%%%%%%%%%%%%%%%%%%%%%%%%%%%%%%%%%%%%%
\subsection{Seiberg-Witten map and emergent metric}

In this section, we give formulas of the Seiberg-Witten map for the noncommutative U(1) gauge theory on the K\"ahler manifold.
We rewrite the field strength and the gauge invariant kinetic term of the noncommutative gauge theory
in a commutative picture by using the Seiberg-Witten map.
The resulting action can be written by an ordinary gauge theory on a deformed background.
Finally, we discuss well-known ambiguity of the Seiberg-Witten map.

Here, we consider the formal power series of $\hbar$ for the gauge field $\hat{A}_{a}$ and the gauge parameter $\hat{\lambda}$.
\begin{align}
	&\hat{A}_{a}=A_{a}+\hbar A_{a}^{(1)}+{\cal O}(\hbar^2),\\
	&\hat{\lambda}=\lambda+\hbar\lambda^{(1)}+{\cal O}(\hbar^2).
\end{align}
The first order of $\hbar$ of the equation (\ref{enbbnsfbjlxg}) is 
\begin{align}
A^{(1)}_{a}(A+d\lambda)-A^{(1)}_{a}(A)={\cal L}_a\lambda^{(1)}-ig^{\bar{\mu}\nu}(\partial_{\bar{\mu}}A_{a}\partial_{\nu}\lambda-\partial_{\bar{\mu}}\lambda \partial_{\nu}A_{a}).
\end{align}
A solution of this equation is given by
\begin{align}
&A_{a}^{(1)}=
\frac{i}{2}g^{\bar{\mu}\nu}(
	A_{\bar{\mu}}(\partial_{\nu}A_{a}+\zeta^{i}_{a}F_{\nu i})
	-(\partial_{\bar{\mu}}A_{a}+\zeta^{i}_{a}F_{\bar{\mu}i})A_{\nu}
),\label{eqnbxlfghbcjgcnnbjkbx}\\
&\lambda^{(1)}=
	\frac{i}{2}g^{\bar{\mu}\nu}(
	A_{\bar{\mu}}\partial_{\nu}\lambda-\partial_{\bar{\mu}}\lambda A_{\nu}
).
\end{align}
To get this results we used holomorphy of the Killing vector and the Killing equations.
The equation (\ref{eqnbxlfghbcjgcnnbjkbx}) is manifestly covariant under the general coordinate transformation 
of the background manifold.

The field strength is expanded as follows:
\begin{align}
\hat{F}_{ab}=&{\cal L}_{a}A_{b}-{\cal L}_{b}A_{a}-if_{ab}^{c}A_{c}\nonumber\\
&+\hbar({\cal L}_{a}A^{(1)}_{b}-{\cal L}_{b}A^{(1)}_{a}-if_{ab}^{c}A^{(1)}_{c}-ig^{\bar{\mu}\nu}\partial_{\bar{\mu}}A_{a}\partial_{\nu}A_{b}+ig^{\bar{\mu}\nu}\partial_{\bar{\mu}}A_{b}\partial_{\nu}A_{a}).
\end{align}
The leading part of order $\hbar^0$ is just the ordinary field strength contracted with the Killing vector.
\begin{align}
{\cal L}_{a}A_{b}-{\cal L}_{b}A_{a}-if_{ab}^{c}A_{c}=\zeta^{i}_{a}\zeta^{j}_{b}F_{ij}=:F_{ab},
\end{align}
where $F_{ij}=\partial_{i}A_{j}-\partial_{j}A_{i}$.
The part of order $\hbar^1$ is
\begin{align}
&{\cal L}_{a}A^{(1)}_{b}
-{\cal L}_{b}A^{(1)}_{a}
-if_{ab}^{c}A^{(1)}_{c}
-ig^{\bar{\mu}\nu}\partial_{\bar{\mu}}A_{a}\partial_{\nu}A_{b}
+ig^{\bar{\mu}\nu}\partial_{\bar{\mu}}A_{b}\partial_{\nu}A_{a}\nonumber\\
&=
ig^{\bar{\mu}\nu}\zeta^{i}_{a}\zeta^{j}_{b}(F_{i\nu}F_{\bar{\mu}j}-F_{j\nu}F_{\bar{\mu}i})
-ig^{\bar{\mu}\nu}(A_{\bar{\mu}}\partial_{\nu}F_{ab}-\partial_{\bar{\mu}}F_{ab}A_{\nu}).
\end{align}
In the commutative picture, the kinetic term of the noncommutative U(1) gauge theory becomes
\begin{align}
S=\frac{1}{4}\int d^{N}zd^{N}\bar{z}\sqrt{g}\left(
	F_{ij}F^{ij}
	-2i\hbar\left(
		F_{\bar{\mu}\nu}g^{\nu\bar{\rho}}F_{\bar{\rho}\sigma}g^{\sigma\bar{\tau}}F_{\bar{\tau}\kappa}g^{\kappa\bar{\mu}}
		+F_{\bar{\mu}\bar{\nu}}g^{\bar{\nu}\rho}F_{\rho\sigma}g^{\sigma\bar{\tau}}F_{\bar{\tau}\kappa}g^{\kappa\bar{\mu}}
		+\frac{1}{4}g^{\bar{\mu}\nu}F_{\bar{\mu}\nu}F_{ij}F^{ij}\right)
	\right),\label{eq:swaction}
\end{align}
where higher corrections are omitted.

The action (\ref{eq:swaction}) can be written by
\begin{align}
S_{g,A}=\frac{1}{4}\int d^{N}zd^{N}\bar{z}\sqrt{g'}g'^{ik}g'^{jl}F_{ij}F_{kl}
\end{align}
with
\begin{align}
 	&g'_{ij}=g_{ij}+\hbar v_{ij},\\
	&v_{\mu\bar{\nu}}=-\frac{i}{2}F_{\mu\bar{\nu}}.
\end{align}
The metric $g'$ is an Hermite metric because the tensor $v$ is Hermite.
For the Levi-Civita connection of the resulting metric $g'$, the Ricci tensor is given by
\begin{align}
R'_{\mu\bar{\nu}}
=&
	R_{\mu\bar{\nu}}
		+\frac{i\hbar}{2}\nabla_{\bar{\nu}}(g^{\bar{\rho}\sigma}\nabla_{\sigma}F_{\mu\bar{\rho}})
		+\frac{i\hbar}{4}\nabla_{\bar{\nu}}(g^{\bar{\rho}\sigma}\nabla_{\bar{\rho}}F_{\sigma\mu})
		-\frac{i\hbar}{4}\nabla_{\sigma}(g^{\bar{\rho}\sigma}\nabla_{\mu}F_{\bar{\nu}\bar{\rho}}),\label{resulnck}
\end{align}
where $R_{\mu\bar{\nu}}$ is the Ricci tensor with respect to the original metric $g$.
The corresponding Ricci scalar is
\begin{align}
R'
=
R
+i\hbar\nabla_{\bar{\mu}}\nabla_{\nu}F^{\bar{\mu}\nu}
+i\hbar F^{\bar{\mu}\nu}R_{\nu\bar{\mu}}.\label{eq:kre}
\end{align}

We discuss deformation of geometric quantities such as the complex structure, the K\"ahler form and the K\"ahler potential.
The complex structure is not changed
\begin{align}
J=
\left(
\begin{array}{cc}
i1_{N}&0\\
0&-i1_{N}
\end{array}
\right).
\end{align}
It can be understood by the fact that the star product is trivial if both functions are (anti-)holomorphic
\begin{align}
f(z)*g(z)=f(z)g(z),\ f(\bar{z})*g(\bar{z})=f(\bar{z})g(\bar{z}).
\end{align}
The corresponding K\"ahler form of the metric $g'$ is given by
\begin{align}
\omega'
=
ig'_{\mu\bar{\nu}}dz^{\mu}\wedge d\bar{z}^{\nu}
=
\omega+\frac{\hbar}{2}F_{\mu\bar{\nu}}dz^{\mu}\wedge d\bar{z}^{\nu},
\end{align}
where $\omega=ig_{\mu\bar{\nu}}dz^{\mu}\wedge d\bar{z}^{\nu}$ is the K\"ahler form of the original metric $g$.
Since the resulting K\"ahler form is not always closed, the manifold $({\cal M},g')$ is not a K\"ahler manifold in general.
If the gauge field configuration satisfies $F_{\mu\nu}=F_{\bar{\mu}\bar{\nu}}=0$, $({\cal M},g')$ is a K\"ahler manifold.
In this case, we can write 
$A_{\mu}=\partial_{\mu}\phi, A_{\bar{\mu}}=\partial_{\bar{\mu}}\phi^{*}$, 
$F_{\mu\bar{\nu}}=\partial_{\mu}A_{\bar{\nu}}-\partial_{\bar{\nu}}A_{\mu}=-\partial_{\mu}\partial_{\bar{\nu}}(\phi-\phi^{*})$ 
where $\phi$ is any functions on the manifold ${\cal M}$. 
For this configuration of the gauge field, the Riemann curvature and the K\"ahler potential for this configuration is given by
\begin{align}
R'_{\mu\bar{\nu}}
=&
	R_{\mu\bar{\nu}}
		+\frac{i\hbar}{2}\partial_{\mu}\partial_{\bar{\nu}}(g^{\bar{\rho}\sigma}F_{\sigma\bar{\rho}}),\\
\Phi'
=&
\Phi+\frac{i\hbar}{2}(\phi-\phi^{*}),
\end{align}
where $\Phi$ is a K\"ahler potential for the metric $g_{\mu\bar{\nu}}$.

Before closing this section, we note that there is well-known ambiguity of the Seiberg-Witten map which is given by
\begin{align}
A_{a}^{(1)}(A;\alpha)=&
A_{a}^{(1)}(A)
+\alpha {\cal L}_{a}(g^{\bar{\nu}\mu}F_{\mu\bar{\nu}}),
\end{align}
where $\alpha$ is an arbitrary constant and $\lambda^{(1)}$ is unchanged.
We can easily show that $\hat{F}_{ab}$ dose not depend on $\alpha$ up to order $\hbar$.
Therefore, if we consider only the gauge field, there is no ambiguity for the gravity side.

%%%%%%%%%%%%%%%%%%%%%%%%%%%%%%%%%%%%%%%%%%%%
\subsection{Relation with contravariant geometry}

We have already provided that some K\"ahler manifolds are the solution of the contravariant gravity 
as explicit examples in the section 2.5.
We assume that the Poisson structure is given by the inverse of the K\"ahler form $\omega$.
Nonzero components of the Poisson tensor are
\begin{align}
\pi^{\mu\bar{\nu}}=iG^{\mu\bar{\nu}}.
\end{align}
On the K\"ahler manifold the Hermite metric is covariantly 
constant with respect to the ordinary Levi-Civita connection, and so is the Poisson tensor.
Then, the contravariant derivative takes a simple form of
\begin{align}
\bar{\nabla}^{i}=\pi^{ij}\nabla_{j}.
\end{align}
In this case, the contravariant Ricci scalar is also simplified as
\begin{align}
\bar{R}(G,\pi)=\pi^{ij}\pi^{mn}R_{jmni}=R(G^{-1}),
\end{align}
where the left hand side is the ordinary Ricci scalar made out of $G_{ij}$.

We focus on the fluctuation of the Poisson tensor, and thus the metric fluctuation is taken to be zero.
The Poisson condition for $\pi^{ij}+\epsilon\rho^{ij}$ can be solved generally by taking
\begin{align}
\rho^{ij}=\pi^{ik}\nabla_{k}{\cal A}^{j}-\pi^{jk}\nabla_{k}{\cal A}^{i},
\end{align}
where ${\cal A}^{i}$ is an arbitrary vector field.
Here, we take
\begin{align}
{\cal A}^{i}=\pi^{ij}A_{j},
\end{align}
by using the U(1) vector field $A_{i}$.
Since the Poisson tensor is covariantly constant, the fluctuation of the Poisson tensor can be written by
\begin{align}
\rho^{ij}=\pi^{ik}\pi^{jl}(\nabla_{k}A_{l}-\nabla_{l}A_{k}).
\end{align}
For this fluctuation of the Poisson tensor the contravariant Ricci scalar can be written by
\begin{align}
\bar{R}(G,\pi+\epsilon\rho)
=&
R(G^{-1})+\epsilon(2G_{il}\pi^{ij}\nabla_{j}\nabla_{k}\rho^{lk}+2\rho^{im}\pi^{nj}R_{jkmn})+{\cal O}(\epsilon^2)\\
=&
R(G^{-1})
+	4\epsilon\left(
		i\nabla_{\bar{\mu}}\nabla_{\nu}\rho^{\bar{\mu}\nu}
		+i\rho^{\bar{\mu}\nu}R_{\bar{\mu}\nu}
		\right)
+{\cal O}(\epsilon^2)\label{eq:cre}
\end{align}
We can see that the contravariant Ricci scalar (\ref{eq:cre}) is the same as the Ricci scalar which 
is obtained by the noncommutative gauge theory on a homogeneous K\"ahler manifold (\ref{eq:kre}) with 
identifying $4\epsilon=\hbar$, $G_{ij}=g_{ij}$, $\rho^{\mu\nu}=-F^{\mu\nu}$ and $\rho^{\bar{\mu}\nu}=F^{\bar{\mu}\nu}$.

%%%%%%%%%%%%%%%%%%%%%%%%%%%%%%%%%%%%%%%%%%%%
\section{Conclusion and outlook}

In this paper, we discussed the emergent gravity from the gauge theories on the Moyal plane 
and on noncommutative homogeneous K\"ahler manifolds from the viewpoint of contravariant gravity.
First, we revisited a formulation of the contravariant gravity. The contravariant Einstein equation and 
the corresponding action were presented and the gauge symmetry of the theory was discussed.
Then, we found some new solutions of the contravariant Einstein equation with or without cosmological constant. 
The linearization around the general background was also examined in detail.
The fluctuations of both metric and Poisson tensor were taken into account.

We started with the discussion of a relation between the contravariant 
gravity and the emergent gravity from a noncommutative U(1) gauge theory on the Moyal plane. 
We focused on the fluctuation of the Poisson tensor 
in the absence of the metric fluctuation in the contravariant gravity side.
Since the Poisson condition in the contravariant geometry can
be interpreted as the Bianchi identity in the gauge theory, 
we could naturally identify the fluctuation of the Poisson tensor $\rho^{ij}$ 
with the gauge field strength $F^{ij}$.
We found that the linearized contravariant Ricci curvature coincides with the Ricci curvature 
of the emergent gravity via the first-order Seiberg-Witten map.

We generalized the above discussion to the case of a gauge theory on a noncommutative homogeneous K\"ahler manifold. 
Since the contravariant geometry contains the metric independently we could extend a discussion on the emergent gravity 
to that from the noncommutative gauge theory on a curved background. 
An explicit expression of the first-order Seiberg-Witten map was provided for the K\"ahler manifold case.
We found that the gauge theory on a noncommutative K\"ahler manifold can be written as an ordinary gauge theory 
on a background deformed from the original K\"ahler geometry. 
The corresponding metric obtained in this procedure turned out to be not necessarily a K\"ahler metric in general. 
We discussed the condition on the gauge field configuration for the obtained geometry to admit a K\"ahler structure.
We argued that the gauge field can be interpreted
as the fluctuation of the Poisson tensor by comparing the curvatures on both sides.
These results indicate that the contravariant gravity is a suitable framework for noncommutative spacetime physics.

We want to comment on our point of view of dynamics of gravity.
Our understanding of the mechanism of the emergent gravity is that the gravity degrees of freedom emerge from the matrix theory.
In our case, we start from a field theory on the noncommutative geometry including ``gravity" as an original theory.
We showed that the U(1) gauge theory on the noncommutative K\"ahler manifold corresponds to the ordinary U(1) gauge theory on the non-K\"ahler background in general, where we have used a mechanism similar to the emergent gravity.
Our result supports that semi-classical approximation of ``gravity" on the noncommutative space is the contravariant gravity.
This is a different approach %from 
but closely related with the emergent gravity \cite{Rivelles,Yang,Steinacker,Steinacker3,YS,Y}.

\vspace{5mm}

There are some interesting directions of this study.
Higher order calculation for the Seiberg-Witten map of the noncommutative gauge theory 
on the curved background is an important part of the emergent gravity.
For the Moyal plane case, the full order explicit formula of the Seiberg-Witten map is given in \cite{BCPVZ,CPZ}
by using BRST symmetry for the noncommutative U(1) gauge theory, see also \cite{Okuyama}.
For the emergent gravity some higher order corrections for the metric are discussed in \cite{Yang,Y3}.
We expect that this geometry can be naturally interpreted in terms of the contravariant geometry.

There is an ambiguity in the Seiberg-Witten map 
due to the non-uniqueness of solutions to the condition (\ref{enbbnsfbjlxg}) \cite{AsakawaKishimoto}.
This ambiguity in the Seiberg-Witten map changes the resulting metric as discussed in \cite{Rivelles2}.
It is interesting to see these results from the contravariant geometry point of view.
See \cite{ADKLW} and references therein for more details of the Seiberg-Witten map.

In this paper, we focused on the ordinary Moyal star product and the star product on K\"ahler manifold which is discussed in \cite{MSSU,Sako}.
Another possibility is to
consider the covariant star product which is given by the symplectic connection on the Fedosov manifold \cite{Vassilevich}.
An interesting question is its relation to the emergent gravity.

It is also interesting to study the emergent gravity on the other curved backgrounds \cite{Y1,Y2}.
Some configurations of curved geometry and the gauge theories on them 
appear from a continuum limit of the matrix theory.
For instance, the fuzzy four sphere is considered in \cite{Steinacker2,SperlingSteinacker1,SperlingSteinacker2}.
In these discussions averaging the Poisson tensor plays a role in recovering spacetime symmetry, 
which is discussed also in \cite{KawaiKawanaSakai}.
The higher spin degrees of freedom are also argued in matrix theory, 
and so it would be an important problem to construct geometric frameworks to handle those degrees of freedom.

%%%%%%%%%%%%%%%%
%%%%%%%%%%%%%%%%

\section*{Acknowledgments}

The authors would like to give thanks to Tsuguhiko Asakawa for valuable comments 
and to Taiki Bessho, Ursula Carow-Watamura, Marc Andre Heller,
Noriaki Ikeda, Goro Ishiki, Tomokazu Kaneko, Takaki Matsumoto, Jeong-Hyuck Park,
Jun-ichi Sakamoto, Yuta Sekiguchi,
Harold Steinacker, Hyun Seok Yang and Kentaroh Yoshida
for fruitful discussions.
YK is supported by Tohoku University Division 
for Interdisciplinary Advanced Research and Education (DIARE).
The work of HM is supported in part by the Iwanami Fujukai Foundation.

%%%%%%%%%%%%%%%%%%%%%%%%%%%%%%%
%%%%%%%%%%%%%%%%%%%%%%%%%%%%%%%

\appendix

%%%%%%%%%%%%%%%%%%%%%%%%%%%%%%%%%%%%%%%%%%%%%%%
\section{Linearization of the contravariant Riemann tensor} \label{sec:linearapp}

We show detail calculations for the formula of the linearization of the contravariant Riemann tensor (\ref{eq:fulllinear}).
The linearized contravariant Levi-Civita connection is
\begin{align}
\bar{\Gamma}^{ij}_{k}(G^{ij}+\epsilon h^{ij},\theta^{ij}+\epsilon\rho^{ij})
=&
	\bar{\Gamma}^{ij}_{k}+\epsilon\bar{\Gamma}'^{ij}_{k}+{\cal O}(\epsilon^2)\\
=&	\bar{\Gamma}^{ij}_{k}\nonumber\\
&	-\epsilon\rho^{im}\Gamma^{j}_{mk}\nonumber\\
&	+\frac{\epsilon}{2}G_{km}
		(\bar{\nabla}^{i}h^{jm}
		+\bar{\nabla}h^{im}
		-\bar{\nabla}^{m}h^{ij})\nonumber\\
&	+\frac{\epsilon}{2}G_{km}
		(G^{in}\nabla_{n}\rho^{mj}
		+G^{jn}\nabla_{n}\rho^{mi}
		+G^{mn}\nabla_{n}\rho^{ij})+{\cal O}(\epsilon^2).
\end{align}
For this connection, we get
\begin{align}
&\bar{R}^{kij}_{l}(G^{ij}+\epsilon h^{ij},\theta^{ij}+\epsilon\rho^{ij})\nonumber\\
&=
	\bar{R}^{kij}_{l}+\epsilon\bar{R}'^{kij}_{l}+{\cal O}(\epsilon^2)\nonumber\\
&=
	\bar{R}^{kij}_{l}\nonumber\\
&~~~
	+\epsilon(\rho^{im}\partial_{m}\bar{\Gamma}^{jk}_{l}
	+\theta^{im}\partial_{m}\bar{\Gamma}'^{jk}_{l}
	-\rho^{jm}\partial_{m}\bar{\Gamma}^{ik}_{l}
	-\theta^{jm}\partial_{m}\bar{\Gamma}'^{ik}_{l}\nonumber\\
&~~~~~~~~~~~~	
	-\partial_{m}\rho^{ij}\bar{\Gamma}^{mk}_{l}
	-\partial_{m}\theta^{ij}\bar{\Gamma}'^{mk}_{l}\nonumber\\
&~~~~~~~~~~~~	
	+\bar{\Gamma}'^{jk}\bar{\Gamma}^{im}_{l}
	+\bar{\Gamma}^{jk}_{m}\bar{\Gamma}'^{im}_{l}
	-\bar{\Gamma}'^{ik}\bar{\Gamma}^{jm}_{l}
	-\bar{\Gamma}^{ik}_{m}\bar{\Gamma}'^{jm}_{l})+{\cal O}(\epsilon^2).
\end{align}
First, we focus on parts of the Riemann tensor including $h^{ij}$.
\begin{align}
\bar{R}'^{kij}_{l}|_{\rho=0}
=&
\frac{1}{2}\theta^{im}\partial_{m}(G_{ln}(\bar{\nabla}^{j}h^{kn}+\bar{\nabla}^{k}h^{jn}-\bar{\nabla}^{n}h^{jk}))-(i\leftrightarrow j)\nonumber\\
&-\frac{1}{2}\partial_{m}\theta^{ij}G_{ln}(\bar{\nabla}^{m}h^{kn}+\bar{\nabla}^{k}h^{mn}-\bar{\nabla}^{n}h^{mk})\nonumber\\
&+\frac{1}{2}G_{mn}(\bar{\nabla}^{j}h^{kn}+\bar{\nabla}^{k}h^{jn}-\bar{\nabla}^{n}h^{jk})\bar{\Gamma}^{im}_{l}-(i\leftrightarrow j)\nonumber\\
&+\frac{1}{2}G_{ln}(\bar{\nabla}^{i}h^{mn}+\bar{\nabla}^{m}h^{in}-\bar{\nabla}^{n}h^{im})\bar{\Gamma}^{jk}_{m}-(i\leftrightarrow j)\nonumber\\
=&
\frac{1}{2}\underbrace{\theta^{im}\partial_{m}G_{ln}}(\bar{\nabla}^{j}h^{kn}+\bar{\nabla}^{k}h^{jn}-\bar{\nabla}^{n}h^{jk})-(i\leftrightarrow j)\nonumber\\
&+\frac{1}{2}G_{ln}(\theta^{im}\partial_{m}\bar{\nabla}^{j}h^{kn}+\theta^{im}\partial_{m}\bar{\nabla}^{k}h^{jn}-\theta^{im}\partial_{m}\bar{\nabla}^{n}h^{jk})-(i\leftrightarrow j)\nonumber\\
&-\frac{1}{2}\partial_{m}\theta^{ij}G_{ln}(\bar{\nabla}^{m}h^{kn}+\bar{\nabla}^{k}h^{mn}-\bar{\nabla}^{n}h^{mk})\nonumber\\
&+\frac{1}{2}\underbrace{G_{mn}\bar{\Gamma}^{im}_{l}}(\bar{\nabla}^{j}h^{kn}+\bar{\nabla}^{k}h^{jn}-\bar{\nabla}^{n}h^{jk})-(i\leftrightarrow j)\nonumber\\
&+\frac{1}{2}G_{ln}(\bar{\nabla}^{i}h^{mn}+\bar{\nabla}^{m}h^{in}-\bar{\nabla}^{n}h^{im})\bar{\Gamma}^{jk}_{m}-(i\leftrightarrow j)\nonumber\\
=&
\frac{1}{2}G_{ln}(\theta^{im}\partial_{m}\bar{\nabla}^{j}h^{kn}+\theta^{im}\partial_{m}\bar{\nabla}^{k}h^{jn}-\theta^{im}\partial_{m}\bar{\nabla}^{n}h^{jk})-(i\leftrightarrow j)\nonumber\\
&-\frac{1}{2}\partial_{m}\theta^{ij}G_{ln}(\bar{\nabla}^{m}h^{kn}+\bar{\nabla}^{k}h^{mn}-\bar{\nabla}^{n}h^{mk})\nonumber\\
&-\frac{1}{2}G_{lm}\bar{\Gamma}^{im}_{n}(\bar{\nabla}^{j}h^{kn}+\bar{\nabla}^{k}h^{jn}-\bar{\nabla}^{n}h^{jk})-(i\leftrightarrow j)\nonumber\\
&+\frac{1}{2}G_{ln}(\bar{\nabla}^{i}h^{mn}+\bar{\nabla}^{m}h^{in}-\bar{\nabla}^{n}h^{im})\bar{\Gamma}^{jk}_{m}-(i\leftrightarrow j)\nonumber\\
=&
\frac{1}{2}G_{ln}(\bar{\nabla}^{i}\bar{\nabla}^{j}h^{kn}+\bar{\nabla}^{i}\bar{\nabla}^{k}h^{jn}-\bar{\nabla}^{i}\bar{\nabla}^{n}h^{jk})-(i\leftrightarrow j)\nonumber\\
=&
\frac{1}{2}G_{ln}(\bar{\nabla}^{i}\bar{\nabla}^{k}h^{jn}-\bar{\nabla}^{i}\bar{\nabla}^{n}h^{jk})-(i\leftrightarrow j)\nonumber\\
&-\frac{1}{2}G_{ln}(\bar{R}^{kij}_{m}h^{mn}+\bar{R}^{nij}_{m}h^{km}),
\end{align}
where in the third equality we used
\begin{align}
\theta^{im}\partial_{m}G_{ln}+G_{nm}\bar{\Gamma}^{im}_{l}=-G_{lm}\bar{\Gamma}^{im}_{n}.
\end{align}

Next, we consider the fluctuation of the Poisson tensor.
\begin{align}
\bar{R}'^{kij}_{l}|_{h=0}
=&
\rho^{im}\partial_{m}\bar{\Gamma}^{jk}_{l}-\rho^{jm}\partial_{m}\bar{\Gamma}^{ik}_{l}\nonumber\\
&+\theta^{im}\partial_{m}\left(-\rho^{jn}\Gamma^{k}_{nl}+\frac{1}{2}G_{ln}(G^{jp}\nabla_{p}\rho^{nk}+G^{kp}\nabla_{p}\rho^{nj}+G^{np}\nabla_{p}\rho^{jk})\right)\nonumber\\
&-\theta^{jm}\partial_{m}\left(-\rho^{in}\Gamma^{k}_{nl}+\frac{1}{2}G_{ln}(G^{ip}\nabla_{p}\rho^{nk}+G^{kp}\nabla_{p}\rho^{ni}+G^{np}\nabla_{p}\rho^{ik})\right)\nonumber\\
&-\partial_{m}\rho^{ij}\bar{\Gamma}^{mk}_{l}\nonumber\\
&-\partial_{m}\theta^{ij}\left(-\rho^{mn}\Gamma^{k}_{nl}+\frac{1}{2}G_{ln}(G^{mp}\nabla_{p}\rho^{nk}+G^{kp}\nabla_{p}\rho^{nm}+G^{np}\nabla_{p}\rho^{mk})\right)\nonumber\\
&+\left(-\rho^{jn}\Gamma^{k}_{nm}+\frac{1}{2}G_{mn}(G^{jp}\nabla_{p}\rho^{nk}+G^{kp}\nabla_{p}\rho^{nj}+G^{np}\nabla_{p}\rho^{jk})\right)\bar{\Gamma}^{im}_{l}\nonumber\\
&-\left(-\rho^{in}\Gamma^{k}_{nm}+\frac{1}{2}G_{mn}(G^{ip}\nabla_{p}\rho^{nk}+G^{kp}\nabla_{p}\rho^{ni}+G^{np}\nabla_{p}\rho^{ik})\right)\bar{\Gamma}^{jm}_{l}\nonumber\\
&+\bar{\Gamma}^{jk}_{m}\left(-\rho^{in}\Gamma^{m}_{nl}+\frac{1}{2}G_{ln}(G^{ip}\nabla_{p}\rho^{nm}+G^{mp}\nabla_{p}\rho^{ni}+G^{np}\nabla_{p}\rho^{im})\right)\nonumber\\
&-\bar{\Gamma}^{ik}_{m}\left(-\rho^{jn}\Gamma^{m}_{nl}+\frac{1}{2}G_{ln}(G^{jp}\nabla_{p}\rho^{nm}+G^{mp}\nabla_{p}\rho^{nj}+G^{np}\nabla_{p}\rho^{jm})\right)
\end{align}
\begin{align}
=&\rho^{im}\partial_{m}\bar{\Gamma}^{jk}_{l}-\rho^{jm}\partial_{m}\bar{\Gamma}^{ik}_{l}\nonumber\\
&-\theta^{im}\partial_{m}\rho^{jn}\Gamma^{k}_{nl}-\rho^{jn}\theta^{im}\partial_{m}\Gamma^{k}_{nl}\nonumber\\
&+\frac{1}{2}\theta^{im}\partial_{m}G_{ln}(\underbrace{G^{jp}\nabla_{p}\rho^{nk}}_{(5)}+\underbrace{G^{kp}\nabla_{p}\rho^{nj}}_{(4)})\nonumber\\
&+\frac{1}{2}G_{ln}(\underline{\underline{\theta^{im}\partial_{m}G^{jp}\nabla_{p}\rho^{nk}}}+\underbrace{\theta^{im}\partial_{m}G^{kp}\nabla_{p}\rho^{nj}}_{(4)})\nonumber\\
&+\underbrace{\frac{1}{2}G_{ln}(G^{jp}\theta^{im}\partial_{m}\nabla_{p}\rho^{nk}+G^{kp}\theta^{im}\partial_{m}\nabla_{p}\rho^{nj}+G^{np}\theta^{im}\partial_{m}\nabla_{p}\rho^{jk})}_{(0)}\nonumber\\
&+\theta^{jm}\partial_{m}\rho^{in}\Gamma^{k}_{nl}+\rho^{in}\theta^{jm}\partial_{m}\Gamma^{k}_{nl}\nonumber\\
&-\frac{1}{2}\theta^{jm}\partial_{m}G_{ln}(\underbrace{G^{ip}\nabla_{p}\rho^{nk}}_{(5')}+\underbrace{G^{kp}\nabla_{p}\rho^{ni}}_{(4')})\nonumber\\
&-\frac{1}{2}G_{ln}(\underline{\underline{\theta^{jm}\partial_{m}G^{ip}\nabla_{p}\rho^{nk}}}+\underbrace{\theta^{jm}\partial_{m}G^{kp}\nabla_{p}\rho^{ni}}_{(4')})\nonumber\\
&-\underbrace{\frac{1}{2}G_{ln}(G^{ip}\theta^{jm}\partial_{m}\nabla_{p}\rho^{nk}+G^{kp}\theta^{jm}\partial_{m}\nabla_{p}\rho^{ni}+G^{np}\theta^{jm}\partial_{m}\nabla_{p}\rho^{ik})}_{(0')}\nonumber\\
&-\partial_{m}\rho^{ij}\bar{\Gamma}^{mk}_{l}\nonumber\\
&+\partial_{m}\theta^{ij}\rho^{mn}\Gamma^{k}_{nl}\nonumber\\
&-\underline{\underline{\frac{1}{2}\partial_{m}\theta^{ij}G_{ln}G^{mp}\nabla_{p}\rho^{nk}}}-\underbrace{\frac{1}{2}\partial_{m}\theta^{ij}G_{ln}G^{kp}\nabla_{p}\rho^{nm}}_{(2)}-\underbrace{\frac{1}{2}\partial_{m}\theta^{ij}\nabla_{l}\rho^{mk}}_{(2)}\nonumber\\
&-\rho^{jn}\Gamma^{k}_{nm}\bar{\Gamma}^{im}_{l}\nonumber\\
&+\frac{1}{2}G_{mn}(\underbrace{G^{jp}\nabla_{p}\rho^{nk}}_{(5)}+\underbrace{G^{kp}\nabla_{p}\rho^{nj}}_{(4)}+\underbrace{G^{np}\nabla_{p}\rho^{jk}}_{(3)})\bar{\Gamma}^{im}_{l}\nonumber\\
&+\rho^{in}\Gamma^{k}_{nm}\bar{\Gamma}^{jm}_{l}\nonumber\\
&-\frac{1}{2}G_{mn}(\underbrace{G^{ip}\nabla_{p}\rho^{nk}}_{(5')}+\underbrace{G^{kp}\nabla_{p}\rho^{ni}}_{(4')}+\underbrace{G^{np}\nabla_{p}\rho^{ik}}_{(3')})\bar{\Gamma}^{jm}_{l}\nonumber\\
&-\rho^{in}\Gamma^{m}_{nl}\bar{\Gamma}^{jk}_{m}\nonumber\\
&+\frac{1}{2}G_{ln}(\underbrace{G^{ip}\nabla_{p}\rho^{nm}}_{(1')}+\underbrace{G^{mp}\nabla_{p}\rho^{ni}}_{(4')}+\underbrace{G^{np}\nabla_{p}\rho^{im}}_{(1')})\bar{\Gamma}^{jk}_{m}\nonumber\\
&+\rho^{jn}\Gamma^{m}_{nl}\bar{\Gamma}^{ik}_{m}\nonumber\\
&-\frac{1}{2}G_{ln}(\underbrace{G^{jp}\nabla_{p}\rho^{nm}}_{(1)}+\underbrace{G^{mp}\nabla_{p}\rho^{nj}}_{(4)}+\underbrace{G^{np}\nabla_{p}\rho^{jm}}_{(1)})\bar{\Gamma}^{ik}_{m}
\end{align}
On the other hand,
\begin{align}
&G^{jp}\bar{\nabla}^{i}\nabla_{p}\rho^{nk}+G^{kp}\bar{\nabla}_{p}\rho^{nj}+G^{np}\bar{\nabla}^{i}\nabla_{p}\rho^{jk}-(i\leftrightarrow j)\nonumber\\
=&
\underbrace{G^{jp}\theta^{im}\partial_{m}\nabla_{p}\rho^{nk}+G^{kp}\theta^{im}\partial_{m}\nabla_{p}\rho^{nj}+G^{np}\theta^{im}\partial_{m}\nabla_{p}\rho^{jk}}_{(0)}+(0')\nonumber\\
&-\underbrace{(G^{jp}\nabla_{p}\rho^{nm}+G^{np}\nabla_{p}\rho^{jm})\bar{\Gamma}^{ik}_{m}}_{(1)}+(1')\nonumber\\
&-\underbrace{(G^{np}\nabla_{p}\rho^{mk}+G^{kp}\nabla_{p}\rho^{nm})\bar{\Gamma}^{ij}_{m}}_{(2)}+(2')\nonumber\\
&-\underbrace{G^{jp}\bar{\Gamma}^{in}_{m}\nabla_{p}\rho^{mk}}_{(5)}+(5')\nonumber\\
&+\underbrace{G^{np}\bar{\Gamma}^{im}_{p}\nabla_{m}\rho^{jk}}_{(3)}+(3')\nonumber\\
&+\underbrace{
G^{kp}\bar{\Gamma}^{im}\nabla_{m}\rho^{nj}-G^{kp}\bar{\Gamma}^{in}_{m}\nabla_{p}\rho^{mj}
-G^{kp}\bar{\Gamma}^{jm}\nabla_{m}\rho^{ni}+G^{kp}\bar{\Gamma}^{jn}_{m}\nabla_{p}\rho^{mi}
}_{(4)+(4')}\nonumber\\
&+\underline{\underline{
G^{jp}\bar{\Gamma}^{im}_{p}\nabla_{m}\rho^{nk}-G^{ip}\bar{\Gamma}^{jm}_{p}\nabla_{m}\rho^{nk}
}}.
\end{align}
We need short calculation for checking (4) and the double underline part.
By summarizing these equation we obtain
\begin{align}
\bar{R}'^{kij}_{l}|_{h=0}
=&\frac{1}{2}G_{ln}(G^{jp}\bar{\nabla}^{i}\nabla_{p}\rho^{nk}+G^{kp}\bar{\nabla}_{p}\rho^{nj}+G^{np}\bar{\nabla}^{i}\nabla_{p}\rho^{jk})-(i\leftrightarrow j)\nonumber\\
&+\underbrace{\rho^{im}\partial_{m}\bar{\Gamma}^{jk}_{l}}_{(A)}-\underbrace{\rho^{jm}\partial_{m}\bar{\Gamma}^{ik}_{l}}_{(B)}\nonumber\\
&-\theta^{im}\partial_{m}\rho^{jn}\Gamma^{k}_{nl}-\underbrace{\rho^{jn}\theta^{im}\partial_{m}\Gamma^{k}_{nl}}_{(B)}\nonumber\\
&+\theta^{jm}\partial_{m}\rho^{in}\Gamma^{k}_{nl}+\underbrace{\rho^{in}\theta^{jm}\partial_{m}\Gamma^{k}_{nl}}_{(A)}\nonumber\\
&-\underline{\partial_{m}\rho^{ij}\bar{\Gamma}^{mk}_{l}}_{(C)}+\partial_{m}\theta^{ij}\rho^{mn}\Gamma^{k}_{nl}\nonumber\\
&-\underbrace{\rho^{jn}\Gamma^{k}_{nm}\bar{\Gamma}^{im}_{l}}_{(B)}
+\underbrace{\rho^{in}\Gamma^{k}_{nm}\bar{\Gamma}^{jm}_{l}}_{(A)}
-\underbrace{\rho^{in}\Gamma^{m}_{nl}\bar{\Gamma}^{jk}_{m}}_{(A)}
+\underbrace{\rho^{jn}\Gamma^{m}_{nl}\bar{\Gamma}^{ik}_{m}}_{(B)}.
\end{align}
Here, each parts with underbraces are
\begin{align}
(A)
=&
\rho^{im}\theta^{pj}R^{k}_{lmp}+\rho^{im}\nabla_{m}K^{jk}_{l}
+\rho^{im}\partial_{m}\theta^{nj}\Gamma^{k}_{nl}-\Gamma^{j}_{mn}K^{nk}_{l}\rho^{im},
\end{align}
\begin{align}
(B)=-(\text{exchange} ~i\leftrightarrow j ~\text{in}~ (A)),
\end{align}
\begin{align}
(C)=\theta^{nm}\partial_{m}\rho^{ij}\Gamma^{k}_{nl}+\partial_{m}\rho^{ij}K^{mk}_{l}.
\end{align}
Finally, we obtain
\begin{align}
\bar{R}'^{kij}_{l}|_{h=0}
=
&	\frac{\epsilon}{2}G_{ln}
		(G^{jm}\bar{\nabla}^{i}\nabla_{m}\rho^{nk}
		+G^{km}\bar{\nabla}^{i}\nabla_{m}\rho^{nj}
		+G^{mn}\bar{\nabla}^{i}\nabla_{m}\rho^{jk})\nonumber\\
&	-\frac{\epsilon}{2}G_{ln}
		(G^{im}\bar{\nabla}^{j}\nabla_{m}\rho^{nk}
		+G^{km}\bar{\nabla}^{j}\nabla_{m}\rho^{ni}
		+G^{mn}\bar{\nabla}^{j}\nabla_{m}\rho^{ik})\nonumber\\
&	-\epsilon\rho^{im}\theta^{jn}R^{k}_{lmn}
	+\epsilon\rho^{jm}\theta^{in}R^{k}_{lmn}\nonumber\\
&	+\epsilon\rho^{im}\nabla_{m}K^{jk}_{l}
	-\epsilon\rho^{jm}\nabla_{m}K^{ik}_{l}
	-\epsilon\nabla_{m}\rho^{ij}K^{mk}_{l}\nonumber\\
&
	-(\rho^{im}\partial_{m}\theta^{jn}+\rho^{jm}\partial_{m}\theta^{ni}+\rho^{nm}\partial_{m}\theta^{ij}+\theta^{im}\partial_{m}\rho^{jn}+\theta^{jm}\partial_{m}\rho^{ni}+\theta^{nm}\partial_{m}\rho^{ij})\Gamma^{k}_{nl}.\label{bnuseobnxhjfbhjx}
\end{align}
The last line (\ref{bnuseobnxhjfbhjx}) vanishes if  the Poisson condition is satisfied for a bi-vector $\theta^{ij}+\epsilon\rho^{ij}$.

%%%%%%%%%%%%%%%%%%%%%%%%%%%%%%%%%%%%%%%%%%%%%%%
\section{Details of curvature on K\"ahler manifold}

We show detail calculations for the formula of the Ricci curvature from the gauge theory 
on the homogeneous K\"ahler manifold (\ref{resulnck}).
Let $\Gamma$ be the ordinary Levi-Civita connection. For any connection $\Gamma_{c}$ 
which has torsion in general, we can write
\begin{align}
\Gamma^{k}_{ij}=\Gamma_{c}{}^{k}_{ij}+T^{k}_{ij},
\end{align}
where $T$ is a contorsion tensor of the connection $\Gamma_{c}$.
For this connection, we get
\begin{align}
R^{k}_{lij}
=&
	R_{c}{}^{k}_{lij}\nonumber\\
&
	+\partial_{I}T^{j}_{jl}-\partial_{j}T^{k}_{il}+T^{m}_{jl}T^{k}_{im}-T^{m}_{il}T^{k}_{jm}\nonumber\\
&
	+\Gamma_{c}{}^{m}_{jl}T^{k}_{im}-\Gamma_{c}{}^{m}_{il}T^{k}_{jk}+\Gamma_{c}{}^{k}_{im}T^{m}_{jl}-\Gamma_{c}{}^{k}_{jm}T^{m}_{il}\nonumber\\
=&
	R_{c}{}^{k}_{lij}\nonumber\\
&
	+\nabla_{(c)}{}_{i}T^{k}_{jl}-\nabla_{(c)}{}_{j}T^{k}_{il}\nonumber\\
&
	+T^{m}_{jl}T^{k}_{im}-T^{m}_{il}T^{k}_{jm}+(T^{m}_{ij}-T^{m}_{ji})T^{k}_{ml},
\end{align}
where $R_{c}{}^{k}_{lij}$ is the Riemann curvature for the connection $\Gamma_{c}$ and $\nabla_{(c)}$ is a covariant derivative with respect to the connection $\Gamma_{c}$.

Let us consider $\Gamma_{c}$ as the Hermitian connection. In this case, 
the nontrivial components of the connection are given by
\begin{align}
\Gamma_{c}{}^{\mu}_{\nu\rho}=g^{\bar{\sigma}\mu}\partial_{\nu}g_{\rho\bar{\sigma}},
\end{align}
where $\mu,\nu,...$ and $\bar{\mu},\bar{\nu},...$ denote indices of complex coordinates.
In this case, the components of the contorsion tensor are 
\begin{align}
T^{\mu}_{\nu\rho}=&\frac{1}{2}g^{\bar{\sigma}\mu}(\partial_{\rho}g_{\nu\bar{\sigma}}-\partial_{\nu}g_{\rho\bar{\sigma}}),\\
T^{\mu}_{\nu\bar{\rho}}=&\frac{1}{2}g^{\bar{\sigma}\mu}(\partial_{\bar{\rho}}g_{\nu\bar{\sigma}}-\partial_{\bar{\sigma}}g_{\nu\bar{\rho}}),\\
T^{\mu}_{\bar{\nu}\rho}=&\frac{1}{2}g^{\bar{\sigma}\mu}(\partial_{\bar{\nu}}g_{\rho\bar{\sigma}}-\partial_{\bar{\sigma}}g_{\rho\bar{\nu}}),\\
T^{\mu}_{\bar{\nu}\bar{\rho}}=&0.
\end{align}
For the Hermite connection, the nontrivial components of the Riemann curvature are
\begin{align}
R_{c}{}^{\mu}_{\nu\bar{\rho}\sigma}=-R_{c}{}^{\mu}_{\nu\sigma\bar{\rho}}=\partial_{\bar{\rho}}(g^{\bar{\kappa}\mu}\partial_{\sigma}g_{\nu\bar{\kappa}}).
\end{align}

So far we consider the general connection.
Here, we compute the $\hbar$ correction of the Riemann curvature of the Levi-Civita connection for the Hermitian metric
\begin{align}
g'_{\mu\bar{\nu}}=&g_{\mu\bar{\nu}}+h_{\mu\bar{\nu}}=g_{\mu\bar{\nu}}-\frac{i\hbar}{2}F_{\mu\bar{\nu}},
\end{align}
where we assume that the background metric $g_{\mu\bar{\nu}}$ is a K\"ahler metric.
In this case, the contorsion tensor for the background metric vanishes, we get
\begin{align}
\Gamma{}^{\mu}_{\nu\rho}=\Gamma_{c}{}^{\mu}_{\nu\rho}=g^{\bar{\sigma}\mu}\partial_{\nu}g_{\rho\bar{\sigma}}.
\end{align}
For the metric $g'$ we get
\begin{align}
T'^{\mu}_{\nu\rho}
=&
-\frac{i\hbar}{4}g^{\bar{\sigma}\mu}\nabla_{\bar{\sigma}}F_{\nu\rho},\label{ghsuirbndxguysrehbfxdjbn1}\\
T'^{\mu}_{\nu\bar{\rho}}
=&
-\frac{i\hbar}{4}g^{\bar{\sigma}\mu}\nabla_{\nu}F_{\bar{\rho}\bar{\sigma}},\\
T'^{\mu}_{\bar{\nu}\rho}
=&
-\frac{i\hbar}{4}g^{\bar{\sigma}\mu}\nabla_{\rho}F_{\bar{\nu}\bar{\sigma}},\\
T'^{\mu}_{\bar{\nu}\bar{\rho}}
=&0.\label{ghsuirbndxguysrehbfxdjbn2}
\end{align}
where the covariant derivative is given by the background connection.
The nontrivial components of the Riemann curvature of the Hermite connection of $g'$ are
\begin{align}
R'_{c}{}^{\mu}_{\nu\bar{\rho}\sigma}
=
-R'_{c}{}^{\mu}_{\nu\sigma\bar{\rho}}
=
\partial_{\bar{\rho}}(g'^{\bar{\kappa}\mu}\partial_{\sigma}g'_{\nu\bar{\kappa}})
=
R{}^{\mu}_{\nu\bar{\rho}\sigma}+\nabla_{\bar{\rho}}(g^{\bar{\kappa}\mu}\nabla_{\sigma}h_{\nu\bar{\kappa}}).\label{ghsuirbndxguysrehbfxdjbn11}
\end{align}
Since $T'$ is order $\hbar$, the total Riemann curvature can be approximated by
\begin{align}
R'^{k}_{lij}
=&
	R'_{c}{}^{k}_{lij}+\nabla{}_{l}T'^{k}_{jl}-\nabla{}_{j}T'^{k}_{il}.
\end{align}
Using (\ref{ghsuirbndxguysrehbfxdjbn1}) to (\ref{ghsuirbndxguysrehbfxdjbn2}) and (\ref{ghsuirbndxguysrehbfxdjbn11}), we obtain
\begin{align}
R'^{\mu}_{\nu\bar{\rho}\sigma}
=&
R^{\mu}_{\nu\bar{\rho}\sigma}
-\frac{i\hbar}{2}\nabla_{\bar{\rho}}(g^{\bar{\kappa}\mu}\nabla_{\sigma}F_{\nu\bar{\kappa}})
-\frac{i\hbar}{4}\nabla_{\bar{\rho}}(g^{\bar{\kappa}\mu}\nabla_{\bar{\kappa}}F_{\rho\nu})
+\frac{i\hbar}{4}\nabla_{\sigma}(g^{\bar{\kappa}\mu}\nabla_{\nu}F_{\bar{\rho}\bar{\kappa}}),\\
R'^{\mu}_{\nu\rho\sigma}
=&
-\frac{i\hbar}{4}\nabla_{\rho}(g^{\bar{\kappa}\mu}\nabla_{\bar{\kappa}}F_{\sigma\nu})
+\frac{i\hbar}{4}\nabla_{\sigma}(g^{\bar{\kappa}\mu}\nabla_{\bar{\kappa}}F_{\rho\nu}),\\
R'^{\mu}_{\nu\bar{\rho}\bar{\sigma}}
=&
-\frac{i\hbar}{4}\nabla_{\bar{\rho}}(g^{\bar{\kappa}\mu}\nabla_{\nu}F_{\bar{\sigma}\bar{\kappa}})
+\frac{i\hbar}{4}\nabla_{\bar{\sigma}}(g^{\bar{\kappa}\mu}\nabla_{\nu}F_{\bar{\rho}\bar{\kappa}}),\\
R'^{\mu}_{\bar{\nu}\rho\sigma}
=&
0,\\
R'^{\mu}_{\bar{\nu}\bar{\rho}\sigma}
=&
-\frac{i\hbar}{4}\nabla_{\bar{\rho}}(g^{\bar{\kappa}\mu}\nabla_{\sigma}F_{\bar{\nu}\bar{\kappa}}),\\
R'^{\mu}_{\bar{\nu}\bar{\rho}\sigma}
=&
0.
\end{align}
The Ricci tensor turns out to be
\begin{align}
R'_{\mu\bar{\nu}}
=&
R'^{\rho}_{\mu\rho\bar{\nu}}+R'^{\bar{\rho}}_{\mu\bar{\rho}\bar{\nu}}\nonumber\\
=&
	R_{\mu\bar{\nu}}
		+\frac{i\hbar}{2}\nabla_{\bar{\nu}}(g^{\bar{\rho}\sigma}\nabla_{\sigma}F_{\mu\bar{\rho}})
		+\frac{i\hbar}{4}\nabla_{\bar{\nu}}(g^{\bar{\rho}\sigma}\nabla_{\bar{\rho}}F_{\sigma\mu})
		-\frac{i\hbar}{4}\nabla_{\sigma}(g^{\bar{\rho}\sigma}\nabla_{\mu}F_{\bar{\nu}\bar{\rho}}).
\end{align}

\section{Poisson geometry and noncommutative emergent gravity}
In the section 3.1, we reviewed the effective metric obtained from the noncommutative gauge theory \cite{Rivelles,Yang}.
The emergent gravity is also discussed in the context of the matrix model \cite{Steinacker,Steinacker3}.
We discuss here the effective metric of a scalar field on generic $n$-dimensional spacetime from the matrix model along with an argument by \cite{KuntnerSteinacker}.
Then, we consider relation between this argument and the contravariant geometry.

Let us consider a simple matrix model action with a scalar field $\Phi$:
\begin{align}
-\text{tr}[X^{a},X^{b}][X_{a},X_{b}]-\text{tr}[X^{a},\Phi][X_{a},\Phi],
\end{align}
where the indices are contracted by a flat metric.
We interpret the matrix $X^a$ as an embedding function $x^a$:
\begin{align}
X^a\sim x^a.
\end{align}
When we formulate the theory using the deformation quantization, 
the matrix commutator is represented by the commutator of the star product $[f,g]_{*}=f*g-g*f$.
In a semi-classical approximation, the star product commutator reduces to the Poisson bracket $[f,g]_{PB}=i\pi^{ij}\partial_if\partial_jg$.
And thus, in the semi-classical approximation, the commutator of matrices is given by 
\begin{align}
[X^a,X^b]\sim [x^a,x^b]_{PB}=i\pi^{ij}(x)\partial_i x^a \partial_j x^b,
\end{align}
where $\pi^{ij}$ is a Poisson tensor and also
\begin{align}
[X^{a},\Phi]_{PB}\sim i\pi^{ij}\partial_i x^a \partial_j\Phi.
\end{align}
In this semi-classical approximation, the action turns out to be
\begin{align}
\int d^{n}x\sqrt{\det\omega}(\pi^{ij}\pi^{kl}g_{ik}g_{jl}+g_{ij}\pi^{ik}\pi^{jl}\partial_{k}\Phi(x)\partial_{l}\Phi(x)),
\end{align}
where $g_{ij}$ is an induced metric given by the embedding function and $\omega$ is the inverse of the Poisson tensor.
From this action, the effective metric of the scalar field is 
\begin{align}
\tilde{G}^{ij}=e^{-\sigma}\pi^{ik}\pi^{jl}g_{kl},
\end{align}
where
\begin{align}
(e^{\sigma})^{\frac{n-2}{2}}=\sqrt{\det g\det \pi}.
\end{align}
The equation of motion of the Poisson tensor is
\begin{align}
\tilde{\nabla}_{k}(e^{\sigma}\tilde{G}^{ki}\tilde{G}^{lj}\omega_{ij})+\tilde{\nabla}_{j}(e^{-\sigma}\eta\pi^{jl})=0,
\end{align}
where $\tilde{\nabla}$ is a covariant derivative of the Levi-Civita connection of the effective metric and $\eta=\frac{1}{4}e^{\sigma}\tilde{G}^{ij}g_{ij}$.
Here, in \cite{KuntnerSteinacker} the following ansatz is taken to solve the equation of motion:
\begin{align}
\tilde{G}_{ij}=g_{ij}.
\end{align}
For generic spacetime dimension, the equation of motion of the Poisson tensor becomes
\begin{align}
e^{-\sigma}=\text{const.},\label{apc1}\\
\tilde{\nabla}^i\omega_{ij}=0.\label{apc2}
\end{align}
There are more detail analysis \cite{KuntnerSteinacker} for the case in four-dimensional spacetime.

For the background which satisfies the equations (\ref{apc1}) and (\ref{apc2}), we can see that the equation of motion of the scalar field can be written in terms of the contravariant Levi-Civita connection on the contravariant geometry: 
\begin{align}
[X^i,[X_i,\Phi]]=\bar{\nabla}_{i}(d_{\pi}\Phi)^i,
\end{align}
where the indices are contracted by the embedding metric.
Furthermore, the contravariant Ricci curvature becomes
\begin{align}
\bar{R}^{ij}=R^{ij},
\end{align}
where $R_{ij}$ is ordinary Ricci tensor.
The typical examples of such backgrounds are the K\"ahler manifolds which is discussed in the section 2.5.
In the emergent gravity, the dimensional reductions to the four-dimensional spacetime and the self-duality of the symplectic form have also been studied \cite{Steinacker,Steinacker3,KuntnerSteinacker}.
These arguments might give deeper understanding for the contravariant gravity.

%%%%%%%%%%%%%%%%%%%%%%%%%%%%%%%%%%%%%%%%%%%%

%%%%%%%%%%%%%%%%%%%%%%%%%%%%%%%%%%%%%%%%%%%%%%%
\end{document}